\documentclass{aa}  

\usepackage{graphicx}
\usepackage{txfonts}
\usepackage[colorlinks=true,linkcolor=blue,urlcolor=blue,citecolor=blue]{hyperref}
\usepackage{natbib}
\usepackage{morefloats}
\usepackage{epstopdf}
\usepackage{ulem}
\usepackage{xspace}
\bibpunct{(}{)}{;}{a}{}{,} %% natbib format for A&A and ApJ

\newcommand{\psr}{\protect PSR~B1259-63/LS2883\xspace} %\protect might be required to use the command in the section titles
\newcommand{\ls}{LS~5039}
\newcommand{\fermi}{{\it Fermi} LAT\xspace}

\newcommand{\nustar}{{\it NuSTAR}\xspace}

\begin{document} 

\title{Non-thermal radiation from a pulsar wind interacting with an inhomogeneous stellar wind}
\author{ 
V. M. de~la~Cita\inst{1}
\and V. Bosch-Ramon\inst{1}
\and X. Paredes-Fortuny\inst{1}
\and D. Khangulyan\inst{2}
\and M. Perucho\inst{3,4}
%%\and M. Rib\'o\inst{1,}\thanks{Serra H\'unter Fellow.}
}

\institute{Departament d'Astronomia i Meteorologia, Institut de Ci\`ences del Cosmos,
Universitat de Barcelona, IEEC-UB, Mart\'{\i} i Franqu\`es 1,
E-08028 Barcelona, Spain
\and Department of Physics, Rikkyo University 3-34-1, Nishi-Ikebukuro, Toshima-ku, Tokyo 171-8501, Japan
\and Departament d'Astronomia i Astrof\'{\i}sica, Universitat de Val\`encia, 
Av.\ Vicent Andr\'es Estell\'es s/n, 46100 Burjassot (Val\`encia), Spain
\and Observatori Astron\`omic, Universitat de Val\`encia, C/ Catedr\`atic Jos\'e Beltran, 2, 46980, Paterna (Val\`encia), Spain 
}

   \date{Received - ; accepted -}
 % \abstract{}{}{}{}{} 
% 5 {} token are mandatory
 
  \abstract
  {Binaries hosting a massive star and a non-accreting pulsar are powerful non-thermal emitters due to the interaction of the pulsar and the stellar wind. The winds of massive stars are thought to be inhomogeneous, which could have an impact on the non-thermal emission.}
  {We study numerically the impact of the presence of inhomogeneities or clumps in the stellar wind on the high-energy non-thermal radiation of high-mass binaries hosting a non-accreting pulsar.}
  {We compute the trajectories and physical properties of the streamlines in the shocked pulsar wind without clumps, with a small clump, and with a large one. This information is used to characterize the injection and the steady state distribution of non-thermal particles accelerated at shocks formed in the pulsar wind. The synchrotron and inverse Compton emission from these non-thermal particles is calculated, accounting also for the effect of gamma-ray absorption through pair creation. A specific study is done for \psr.}
  {When stellar wind clumps perturb the two-wind interaction region, the associated non-thermal radiation in the X-ray band, of synchrotron origin, and in the GeV--TeV band, of inverse Compton origin, is affected by several equally important effects: (i) strong changes in the the plasma velocity direction that result in Doppler boosting factor variations, (ii) strengthening of the magnetic field that mainly enhances the synchrotron radiation, (iii) strengthening of the pulsar wind kinetic energy dissipation at the shock, potentially available for particle acceleration, and (iv) changes in the rate of adiabatic losses that affect the lower energy part of the non-thermal particle population. The radiation above 100~GeV detected, presumably, during the post-periastron crossing of the Be star disc in \psr, can be roughly reproduced assuming that the crossing of the disc is modeled as the encounter with a large inhomogeneity.}
  {Because of the likely diverse nature of clumps in the stellar wind, and hydrodynamical instabilities, the non-thermal radiation of high-mass binaries with a non-accreting pulsar is expected to be boosted somewhat chaotically, and present different superimposed variability patterns. Some of the observed variability in gamma rays from \psr is qualitatively reproduced by our calculations.}

   \maketitle
   
%%%%%%%%%%%%%%%%%%%%%%%%%
\section{Introduction}\label{sec:intro}
%%%%%%%%%%%%%%%%%%%%%%%%%

In binaries hosting a high-mass star and a young non-accreting pulsar, strong interaction between the relativistic pulsar wind and the stellar wind is expected. In these wind collisions, efficient particle acceleration and non-thermal emission can take place \citep[][]{MarTre81,tak94}, which would be behind the emission observed from radio to gamma rays in some of these objects, like {\psr} \citep[e.g.][]{aaa05,can14}. Given that this emission, or at least a significant fraction of it, is expected to originate in the region where the winds collide, a proper characterization of the stellar and the pulsar wind is needed to understand the involved physical processes.

Density inhomogeneities, or clumps, are thought to be present in the stellar winds of early-type stars \citep{LucSol70}. The hydrodynamical and radiative consequences of the presence of clumps were studied analytically in \cite{Bos13}, and relativistic, axisymmetric, hydrodynamical (RHD) simulations were carried out by \cite{pbp15} to study in more detail the impact of different types of clumps on the two-wind interaction region. It was found that clumps can noticeably affect the shape, size, and the stability of the interaction structure, and the variability patterns of the radiation coming from there. It was also proposed that wind inhomogeneities could be responsible of the GeV flare in {\psr} \citep{can14}, and may play also an important role in the X-ray activity of some binaries \citep[e.g., see the discussion in][and references therein]{Bos13}. However, the impact of the presence of inhomogeneities in the stellar wind in the high-energy emission has not been accurately studied yet.
 
In this work, we compute for the first time the synchrotron and IC emission produced by the interaction of an inhomogeneous\footnote{Inhomogeneous means here clumpy, whereas homogeneous means non clumpy.} stellar wind and a pulsar wind based on hydrodynamic simulations, obtaining the spectral energy distributions (SED) and maps of the emitting region. To characterize the impact of wind inhomogeneities on the non-thermal radiation, we have used the flow information obtained from the RHD simulations done by \cite{pbp15}. From the flow hydrodynamical quantities, we have obtained a number of streamlines, characterizing the fluid of interest in the form of several 1D structures from which we compute, following the method described in \cite{bp16}, the synchrotron and the inverse Compton (IC) radiation. As the stellar photon field is very dense, gamma-ray absorption due to electron-positron pair creation has been taken into account. An approach similar to the one adopted here was followed by \cite{dlf15} in a broad study of the non-thermal emission of high-mass binaries hosting a non-accreting pulsar, although in that case the stellar wind was assumed to be homogeneous.

A region of a size similar to the star-pulsar separation distance is considered. The reason is three-fold:
(i) in this paper, we are mostly concerned with the main radiation features resulting from the interaction of a clump with the two-wind collision structure; (ii) we are interested in the highest energies, which are expected to be produced on the binary scales \citep[see nevertheless][]{zba13}; (iii) for reasons explained in \cite{pbp15}, simulation results were limited to these scales. Since radio emission is expected to be produced far from the binary system  \citep[e.g.][]{dub06*b,bos11}, the focus here is put on X-rays and gamma rays.

Regarding the most recent simulations of stellar and pulsar winds collisions, few important differences from our work are to be mentioned. The simulations in \cite{dlf15} are three-dimensional (3D), whereas here the hydrodynamical results are taken from the simulations of \cite{pbp15}, carried out with axisymmetry (2D). In addition, the grid was significantly larger in \cite{dlf15} compared with \cite{pbp15}. Regarding \cite{bbp15}, the 3D simulations included orbital motion, which proved to be important beyond few star-pulsar separation distances. That said, we remark that the relatively small size of the computational grid in the present work allows one to neglect orbital motion, although instability development may be slower in our 2D simulations (see Sect.~\ref{sec:hydro}), and 3D calculations may yield remarkable quantitative differences in general. We note as well that the dynamical role of the magnetic field was not included in \cite{pbp15}. So far, only \cite{bkk12} has included the magnetic field when computing the two-wind interaction structure in the context of binary systems, whereas several works have studied in 1D, 2D and 3D the magnetohydrodynamics, and in some cases the radiation, for isolated pulsars interacting with the environment in the relativistic regime \citep[e.g.][and references therein]{bad05,vda08,oda14,pkk14,mlv15,YooHei16}.

The paper is distributed as follows: in Sect.~\ref{sec:hydro}, we describe the hydrodynamic setup and results; in Sect.~\ref{sec:radiative}, we present the radiative computation results for the different cases of study together with an application to the gamma-ray binary \psr; finally, in Sects.~\ref{sec:conclusions} and \ref{sec:discussion}, we sum up our conclusions and discuss the results.

%%%%%%%%%%%%%%%%%%%%%%%%%
\section{Hydrodynamics}\label{sec:hydro} %
%%%%%%%%%%%%%%%%%%%%%%%%%

%The hydrodynamical results from these simulations are used here to compute the physical variables of the flow along the trajectories followed by the injected pulsar wind. Then, using that information, the radiative output from the shocked pulsar wind, the dominant energy source in the region of interest, is computed.

In this section, we summarize the numerical simulations from which the hydrodynamical information is obtained. In addition, we also briefly explain how the pulsar wind is characterised as a set of streamlines. For further details on both the hydrodynamical simulations and the mathematical procedure to obtain the streamlines, we refer to \cite{pbp15} and \cite{bp16} (Appendix A), respectively. The code used in these simulations is a finite-volume, high-resolution shock-capturing scheme that solves the equations of relativistic hydrodynamics in conservation form. This code is an upgrade of that described in \cite{mmf97}, parallelized using OMP directives \citep{pmh05}. The numerical fluxes at cell boundaries are computed using an approximate Riemann solver that uses the complete characteristic information contained in the Riemann problems between adjacent cells \citep{DonMar96}. It is based on the spectral decomposition of the Jacobian matrices of the relativistic system of equations derived in \cite{fim94}, and uses analytical expressions for the left eigenvectors \citep{dfi98}. The spatial accuracy of the algorithm is improved up to third order by means
of a conservative monotonic parabolic reconstruction of the pressure, proper rest-mass density and the spatial components of the fluid four-velocity (PPM, see \citealt{ColWoo84}, and \citealt{MarMul15}). Integration in time is done simultaneously in both spatial directions using a multi-step total-variation-diminishing (TVD) Runge-Kutta (RK) method developed by \cite{ShuOsh88}, which provides third order in time. The simulations were run in a workstation with two Intel(R) Xeon(R) CPU E5-2643 processors (3.30 GHz, $4 \times 2$ cores, with two threads for each core) and four modules of 4096 MB of memory (DDR3 at 1600 MHz).

\cite{pbp15} performed axisymmetric RHD simulations of the interaction of a relativistic pulsar wind and an inhomogeneous stellar wind. We simulated first a stellar wind without clumps until a steady state of the two-wind interaction region was achieved. Then, a spherical inhomogeneity centered at the axis between the two stars was introduced.
%Then, an inhomogeneity characterized by a single clump centered at the axis joining the two stars was introduced, and its effect on the two-wind interaction structure was followed until reaching again a steady state. 

An ideal gas with a constant adiabatic index of $\hat\gamma = 1.444$, between a relativistic and a non-relativistic index, was adopted. The magnetic field was assumed to be dynamically negligible. The physical size of the domain is $r \in \lbrack 0, l_r \rbrack$ with $l_r = 2.4\times10^{12}$~cm, and $z \in \lbrack 0, l_z \rbrack$ with $l_z = 4.0\times10^{12}$~cm. The star is located outside the simulated grid at $(r_*, z_*) = (0, 4.8\times10^{12})$~cm, and its spherical wind is injected as a boundary condition at the top of the grid. The pulsar is placed inside the grid at  $(r_{\rm p}, z_{\rm p}) = (0, 4\times10^{11})$~cm, and its spherical wind is injected at a radius of $2.4\times10^{11}$~cm (15 cells). The star-pulsar separation is $d=4.4\times10^{12}$~cm. The lower and right boundaries of the grid were set to outflow, whereas the left boundary was set to reflection. The selected physical parameters for the stellar wind at a distance $r = 8\times10^{10}$~cm with respect to the star centre were:
the mass-loss rate $\dot{M}=10^{-7}~{\rm M_\odot~yr^{-1}}$, 
the stellar wind radial velocity $v_{\rm sw}= 3000~{\rm km~s^{-1}}$,
and the specific internal energy $\epsilon_{\rm sw} = 1.8\times10^{15}~{\rm erg~g^{-1}}$;
the derived stellar wind density is $\rho_{\rm sw} = 2.68\times10^{-13}~{\rm g~cm^{-3}}$.
Similarly, the chosen physical parameters for the pulsar wind at a distance $r = 8\times10^{10}$~cm with respect to the pulsar centre were:
the Lorentz factor $\Gamma=5$,
the specific internal energy $\epsilon_{\rm pw} = 9.0\times10^{19}~{\rm erg~g^{-1}}$,
and the pulsar wind density $\rho_{\rm pw} = 2\times10^{-19}~{\rm g~cm^{-3}}$; the derived total pulsar wind luminosity is $L_{\rm p} = 10^{37}~{\rm erg~s^{-1}}$, and the pulsar-to-stellar wind momentum rate ratio $\eta \approx 0.2$.
The two wind parameters are summarized in Table~\ref{rhd_param}.

The simulation adopted resolution was modest, $150$ and $250$ cells in the radial and the vertical directions, respectively. This resolution is high enough to get the main dynamical features of the two-wind interaction structure, but low enough to avoid a too disruptive instability growth, as explained in \cite{pbp15} \citep[see also][]{phm04}. As noted in that work, the fast instability growth in the two-wind collision region is physical \citep[see also][]{bbk12,bbp15}, although the presence of a singularity in the radial coordinate may introduce additional numerical perturbations to the colliding structure. A much larger grid should have allowed the growing instabilities to leave the computational domain without disrupting the simulation, although some trials indicated that even under the same resolution 2-3 times larger grids eventually also led to simulation disruption, filling the whole grid with shocked flow. Therefore, being the goal in \cite{pbp15} (and here) to carry out a preliminary analysis of the problem, the choice adopted was a modest resolution to keep perturbation growth under control, and a relatively small grid size, allowing the simulation to reach a quasi-steady state solution for the case of the pulsar-star wind interaction without clumps.

%The initial conditions of the simulation were set as it follows: first, the grid was divided in two halves, accounting for the location where the on-axis wind ram pressures of both winds become equal ($z\sim1.8\times10^{12}$~cm), then the upper and lower halves were filled with the solution of the spherical Bernoulli equations, for the stellar and pulsar wind, respectively. Same equations were used to compute the physical values at injection for both the stellar and pulsar wind.

Once the (quasi)-steady state was reached, an inhomogeneity was introduced to the stellar wind. The inhomogeneous wind was thus characterised by a single clump placed at  $(r, z) = (0, 2.6\times10^{12})$~cm and parametrized by its radius $R_{\rm c}$ and its density contrast $\chi$ with respect to the density value at the location where it was introduced. A thorough description of the simulations and their results is given in \cite{pbp15}. Here, two cases among those considered in that work are studied: (i) a clump with $\chi=10$ and $R_{\rm c} = 8\times10^{10}$~cm; and (ii) a clump with $\chi=10$ and $R_{\rm c} = 4\times10^{11}$~cm.

\begin{table}
\caption{Stellar and pulsar parameters.}
\label{rhd_param}
\centering
\begin{tabular}{c c c}
\hline\hline
Parameter  & Stellar wind   & Pulsar wind \\
\hline
$v$         & $3\times 10^8$~cm~s$^{-1}$            & $2.94\times 10^{10}$~cm~s$^{-1}$      \\
$\epsilon$  & $1.8\times10^{15}~{\rm erg~g^{-1}}$   & $9\times10^{19}~{\rm erg~g^{-1}}$     \\
$\rho$      & $2.68\times10^{-13}~{\rm g~cm^{-3}}$  & $2\times10^{-19}~{\rm g~cm^{-3}}$  \\
$(r_{\rm */p},z_{\rm */p})$ & $(0,4.8\times10^{12}~{\rm cm})$       & $(0,4\times10^{11}~{\rm cm})$         \\
\hline
\end{tabular}
\tablefoot{
Wind velocity $v$, specific internal energy $\epsilon$, and
density $\rho$ at a distance $r = 8\times10^{10}$~cm with respect to the star/pulsar centres, located at $(r_{\rm */p},z_{\rm */p})$.
}
\end{table}

After obtaining the hydrodynamical information, and prior to the radiative calculations, one has to compute the streamlines of the pulsar wind. A streamline is defined as the trajectory followed by a fluid element in a steady flow, being the steady flow assumption approximately valid taking into account that the dynamical timescale of the two-wind interaction structure is $\sim l_z/v_{\rm w}$, whereas the one for the pulsar wind is $\sim l_z/c$ \citep[see][]{bp16}.

We computed the streamlines starting from a distance $2.4\times10^{11}$~cm from the pulsar centre.  The magnetic field, $B$ in the laboratory frame and $B^{\prime}$ in the fluid frame, is computed under the assumptions that the Poynting flux is a fraction $\chi_B$ of the matter energy flux, that $B$ is frozen into the plasma under ideal MHD conditions, and that it is at injection the toroidal magnetic field of the unshocked pulsar wind, perpendicular to the motion of the flow (as e.g. in \citealt{dlf15}): $B=\Gamma B^{\prime}$. The evolution of $B^{\prime}$ is obtained then from:
\begin{equation}
B^{\prime} = B_0^{\prime} \sqrt{\frac{\rho v_0 \Gamma_0}{\rho_0 v \Gamma}}\,,
\end{equation}
denoting the subscript 0 the origin of the streamline, i.e., the pulsar wind injection surface, and being $\rho$ the density, $v$ the three-velocity modulus, and $\Gamma$ the Lorentz factor. The initial magnetic field at the starting point of each streamline in the flow frame is computed as
\begin{equation}
\frac{B_0^{\prime 2}}{4\pi} = \chi_B \rho_0 h_0 cv_0\label{chi}\,,
\end{equation}
where $h=1+\hat\gamma\epsilon_{\rm pw}/c^2$ is the specific enthalpy. 
Our $\chi_B$-prescription for $B_0^\prime$ (or $B_0$) is slightly different from the $\sigma$-prescription  in \cite{KenCor84}, although $\chi_B$ and $\sigma$ almost coincide for the values of $\Gamma_0$ and $\chi_B$ adopted in this work.
We note that $\chi_B$ cannot be too large (formally $\chi_B\ll 1$; see Sect.~\ref{ntemit}) because this would collide with the assumption of a dynamically negligible magnetic field made when using a purely RHD code.

\begin{figure*}
\centering
\resizebox{0.95\hsize}{!}{\includegraphics{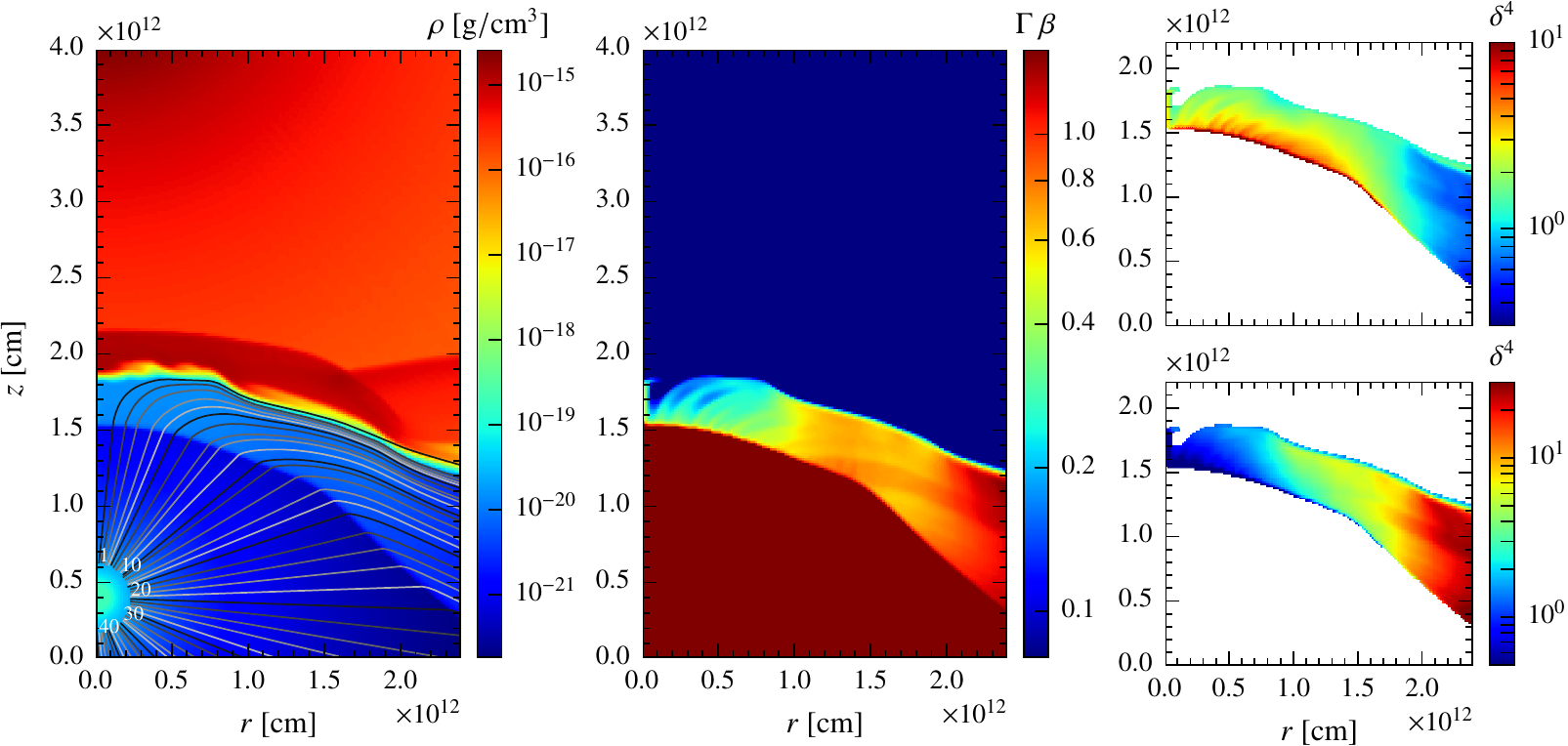}}
\caption{Left panel: density distribution by colour
at time $t = 5.8\times10^{4}$~s in the (quasi)-steady state. 
The star is located at $(r_*, z_*) = (0, 4.8\times10^{12})$~cm and the pulsar wind is injected at a distance of $2.4\times10^{11}$~cm with respect the pulsar centre at $(r_{\rm p}, z_{\rm p}) = (0, 4\times10^{11})$~cm. The grey lines show the obtained streamlines describing the trajectories of the pulsar wind fluid cells; the grey scale and the numbers are only for visualization purposes. Central panel: distribution by colour of the module of the three-velocity at time $t = 5.8\times10^{4}$~s
in the (quasi)-steady state. Right panel: distribution by colour of the Doppler boosting enhancement ($\delta^4$) for the emission produced in the shocked pulsar wind, as seen from $45^\circ$ (top) and $135^\circ$ (bottom) from the pulsar-star axis, at time $t = 5.8\times10^{4}$~s in the (quasi)-steady state.}
\label{steady}
\end{figure*}

\begin{figure*}
\centering
\resizebox{0.95\hsize}{!}{\includegraphics{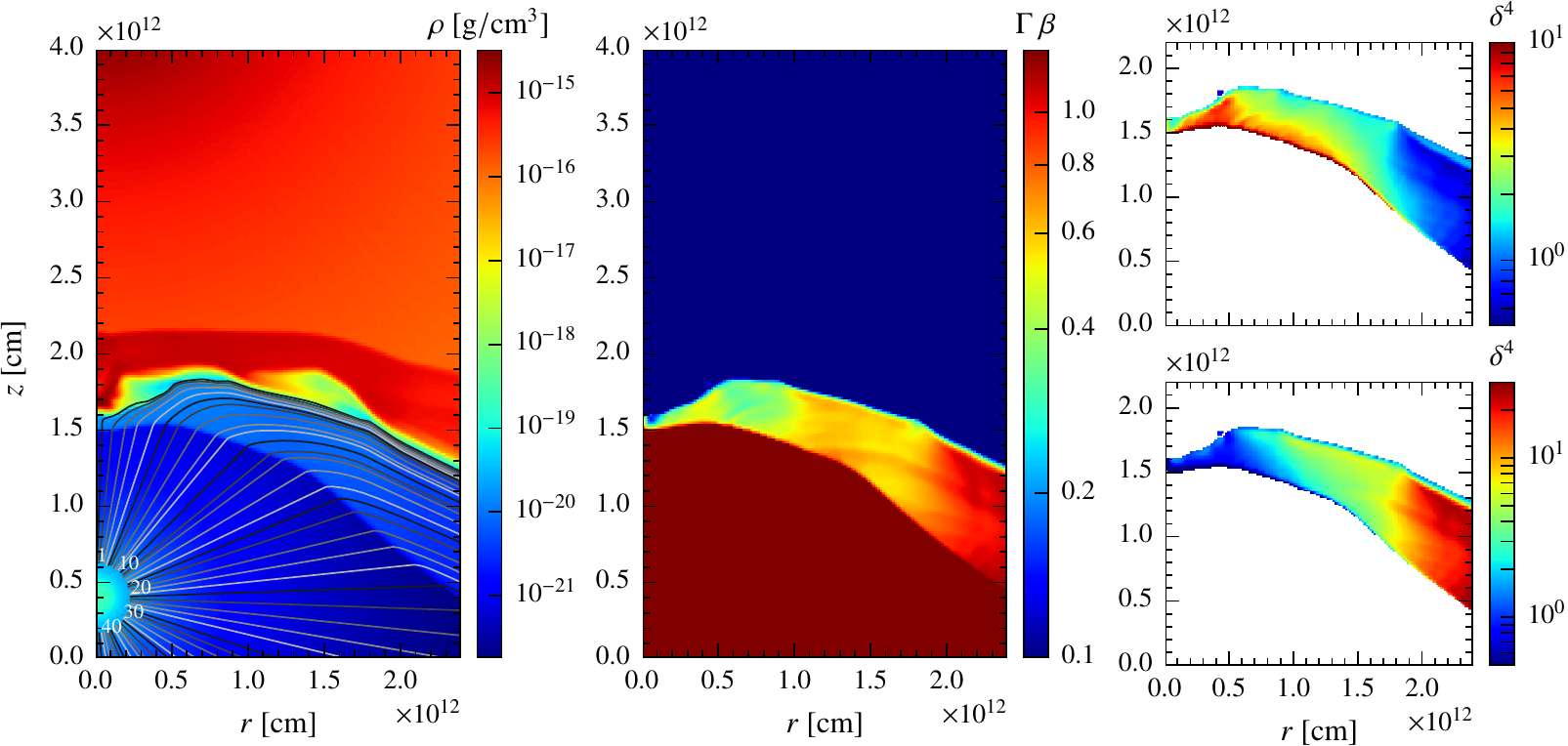}}
\caption{Left panel: density distribution by colour at time $t = 0.4\times10^{4}$~s (measured from the steady case) considering an inhomogeneous stellar wind with $\chi=10$ and $R_{\rm c} = 8\times10^{10}$~cm. The remaining plot properties are the same as those of Fig.~\ref{steady}. Central panel: module of the three-velocity. Right panel: Doppler boosting enhancement as seen from $45^\circ$ (top) and $135^\circ$ (bottom)}
\label{Chi10Rc1}
\end{figure*}

\begin{figure*}
\centering
\resizebox{0.95\hsize}{!}{\includegraphics{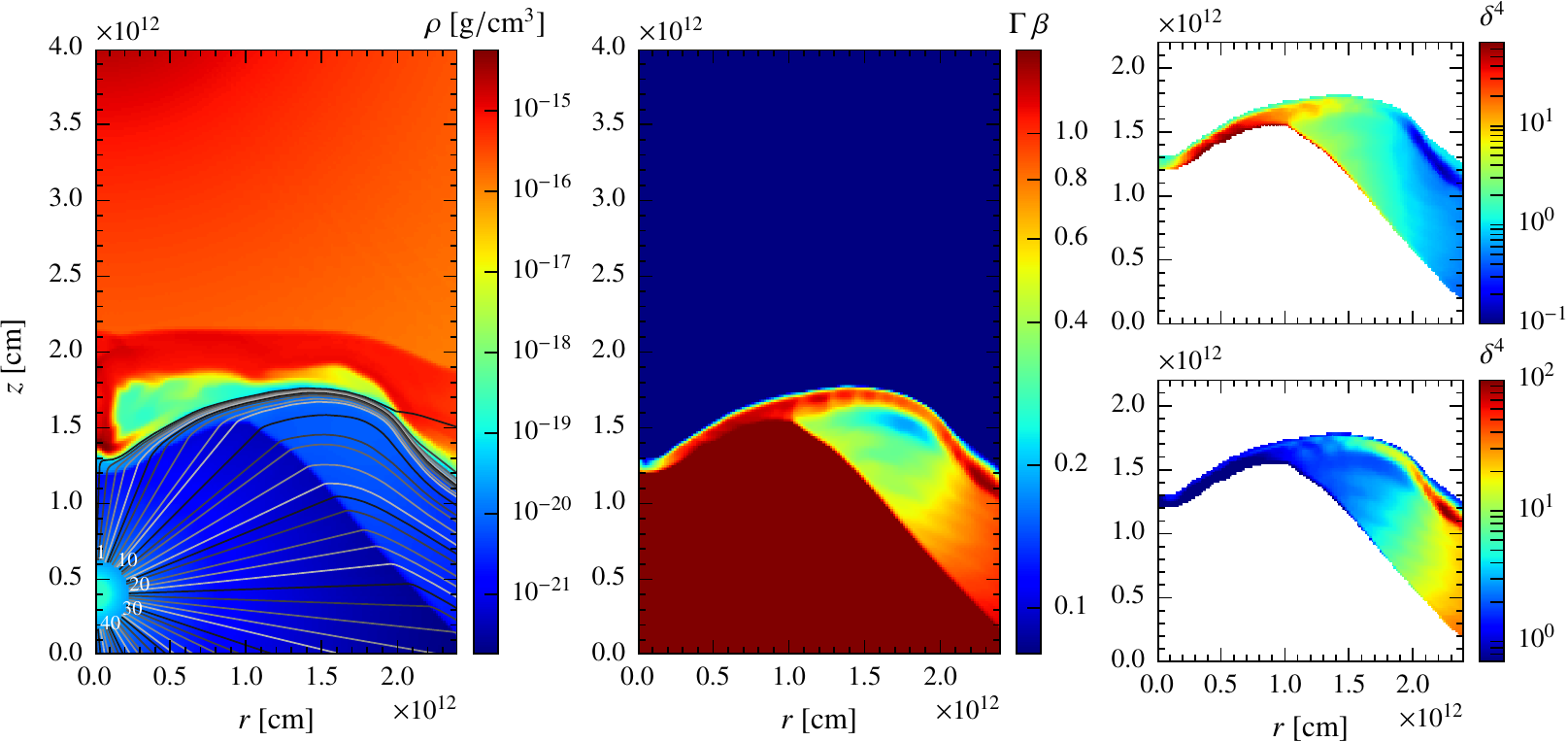}}
\caption{Left panel: density distribution by colour at time $t = 1.1\times10^{4}$~s (measured from the steady case) considering an inhomogeneous stellar wind with $\chi=10$ and $R_{\rm c} = 4\times10^{11}$~cm. The remaining plot properties are the same as those of Fig.~\ref{steady}. Central panel: module of the three-velocity. Right panel: Doppler boosting enhancement as seen from $45^\circ$ (top) and $135^\circ$ (bottom)}
\label{Chi10Rc5}
\end{figure*}

Left panel of Fig.~\ref{steady} shows the computed streamlines plotted on top of the density map for the simulation steady state. Central panel of Fig.~\ref{steady} illustrates the re-acceleration of the shocked pulsar wind as it is advected along the shock. In the right panel of Fig.~\ref{steady} one can see the effect of Doppler boosting quantified by a factor $\delta^4$ for the viewing angles with respect to the pulsar-star axis $\phi=45^\circ$ and $135^\circ$\footnote{An observer with $\phi=0^\circ$ would be looking along the star-pulsar axis.}. The same is shown for the two cases with wind inhomogeneity: Fig.~\ref{Chi10Rc1}, for the clump with $\chi=10$ and $R_{\rm c} = 8\times10^{10}$~cm; and Fig.~\ref{Chi10Rc5}, for the clump with $\chi=10$ and $R_{\rm c} = 4\times10^{11}$~cm.

%%%%%%%%%%%%%%%%%%%%%%%%%
\section{Radiation}\label{sec:radiative}
%%%%%%%%%%%%%%%%%%%%%%%%%
\subsection{The non-thermal emitter}\label{ntemit}

From the streamline information obtained from the RHD simulations, one can derive the characteristics of injection, cooling and radiation of the non-thermal particles as they are advected by the shocked pulsar wind. Each streamline is divided in 200 cells and from each cell we have a set of parameters: position and velocity information, pressure ($P$), density ($\rho$), streamline effective section ($S$),  magnetic field in the flow frame ($B^\prime$), and the flow velocity divergence ($\nabla(\Gamma\overrightarrow{v})$, for the computation of adiabatic losses; see \citealt{bp16}). A tracer accounting for wind mixing, ranging from 0 ($100\%$ pulsar material) to 1 ($100\%$ stellar wind), is provided by the RHD simulations.

In this work, we consider that the non-thermal emitter is restricted to the shocked pulsar wind, i.e., particles are accelerated at the pulsar wind termination shock, although the unshocked pulsar wind may also be an efficient gamma-ray emitter for certain values of the wind Lorentz factor \citep[e.g.][]{AhaBog03,kha07}. For each cell of each streamline, we compute whether there is non-thermal particle injection and the luminosity injected in the form of these particles. The following procedure is applied along each streamline: starting from the line origin, one finds cells where the internal energy increases (and the flow velocity decreases). Non-thermal particles with total energy corresponding to a fixed fraction $\eta_{\rm NT}$ of the internal energy increase are injected in these cells. 
Particle acceleration in  pulsar wind termination shocks is not yet well understood. Our main aim here however is just to show general trends in the radiation due to the presence of clumps. Therefore, we follow a purely phenomenological approach, and assume that the injected particles follow a power-law distribution in energy, with a typical index of $-2$, and two exponential cutoffs at high ($E_{\rm max}$) and low ($E_{\rm min}$) energies. The value of $E_{\rm min}$ is fixed to 1~MeV, and $E_{\rm max}$ is derived adopting an acceleration rate $\sim 0.1\,qB^\prime c$, typical for efficient accelerators.
A value for the fraction $\eta_{\rm NT}$ of $0.1$ is taken, but all the results scale linearly; presently, this quantity cannot be derived from first principles. Note that, as discussed by \cite{dlf15}, a large value of $\eta_{\rm NT}$ would not be consistent if losses of energy (and momentum; see Sect.~\ref{sec:discussion}) were significant in the emitting flow, affecting the flow dynamics. In any case, a formal limitation for $\eta_{\rm NT}$ would be a value such that the energy losses of the non-thermal particles in the laboratory frame (LF) should be well below the energy rate of the pulsar wind (i.e. $L_{\rm p}$; in fact, $\eta_{\rm NT}=0.1$ implies radiation losses of a few \% of $L_{\rm p}$; see Sect.~\ref{resu}).

Once the injection of non-thermal particles is characterised, one computes the energy evolution and spatial propagation of particles along the streamlines until they leave the grid, which eventually leads to a steady state. Then, the synchrotron and IC emission for each cell, for all the streamlines, are computed in the fluid frame (FF), and appropriately transformed afterwards to the observer frame multiplying the photon energies and fluxes by $\delta=1/\Gamma(1-v\cos[\phi_{\rm obs}])$ and $\delta^4$, respectively, where $\phi_{\rm obs}$ is the angle between the flow direction of motion and the line of sight in the LF. A detailed description of the applied method is described in \cite{bp16}, Appendix B. 

In the present scenario, the gamma-ray absorption due to electron-positron pair creation in the stellar photon field cannot be neglected and is taken account. To compute gamma-ray absorption, we have adopted the cross section from \cite{GouSch67}, and assumed that the star is far enough to be point-like \citep[see, e.g.,][]{dub06,BosKha09}. The angle between the direction of the stellar photons, which come from the star centre, and the observer line of sight, is taken into account for both pair production and IC scattering \citep[see, e.g.][for the latter]{AhaAto81,kak14}. With all this information one can build the total spectral energy distributions (SED), and the emission maps of the different radiation channels. To keep the calculations manageable, emission of secondary particles from gamma-ray absorption (i) and IC cascading (ii) has not been considered, but these processes could be indeed important in close binaries, mainly increasing the X-ray fluxes (i; high ambient magnetic field), or enhancing the effective transparency of the system (ii; low ambient magnetic field) \citep[e.g.][]{SieBed05,aak06,SieTor07,bka08,cmd10,BosKha11}. 

The non-thermal emitter is a 3D structure, whereas the hydrodynamic simulations are 2D. Therefore, the streamlines have to be distributed in azimuthal angle around the pulsar-star axis to properly account for IC,  gamma-ray absorption, and Doppler boosting for an observer looking from a certain direction. To do so, an azimuthal random position has been assigned to each streamline cell conserving the values of the $r$- and $z$-components for position and velocity. This transformation, only concerned with the observer direction, affects just the radiative part of the code; the particle energy distribution in each cell is determined by axisymmetric processes and remains unaffected.

The companion star is by far the dominant source of target IC photons. The stellar spectrum has been assumed to be typical for a O-type star: a black body with a temperature and luminosity of $T_\star = 4\times 10^4$~K and $L_\star = 10^{39}$~erg~s$^{-1}$, respectively. The high stellar luminosity allow us to neglect the radiation field produced in the emitter itself, in particular the role of synchrotron self-Compton, which is a safe assumption as long as $L_\star\gg L_{\rm p}/\eta$. The magnetic field is set as specified in Eq.~(\ref{chi}) using two values for $\chi_B$: $10^{-3}$ and  $0.1$, illustrative of a low and a high magnetic field case, respectively. Typically, pulsar winds at termination are considered to be strongly dominated by their kinetic energy \citep[e.g.][]{KenCor84,bkk12,abk12}, although some models predict a high magnetization up to the termination shock, where the magnetic field would efficiently dissipate \citep[e.g.][]{LyuKir01}. We note that for $\chi_B=0.1$ the condition of a negligible magnetic field becomes only marginally fulfilled.

\subsection{Results}\label{resu}
\begin{table}
\centering
\begin{tabular}{c c}
\hline\hline
Parameter & Set of values\\
\hline
Observation angle $\phi$	& 45, 90, $135^\circ$ \\
Fraction $\chi_B$ 			& $10^{-3}, 0.1$\\
\hline
\end{tabular}
\caption{Set of parameters for the three scenarios considered.}
\label{tab:freepara}
\end{table}

\begin{table*}
\caption{Values of the integrated emission in different bands for the case of weak magnetization, $\chi_{\rm B}=10^{-3}$, given in erg~s$^{-1}$; and difference imposed by clumps to the homogeneous case, given in percents.}
\label{tab:bands}
\centering
\resizebox{\textwidth}{!}{%
\begin{tabular}{cccccccccccc}
% BEGIN RECEIVE ORGTBL emission
\hline
\hline
Scenario & $\phi$ & $1-10$ keV & diff., \% & $10-100$ keV & diff., \% & $0.1-100$ MeV & diff., \% & $0.1-100$ GeV & diff., \% & $0.1-100$ TeV & diff., \% \\
\hline
 & 45 & $1.25\times10^{33}$ &  & $4.51\times10^{33}$ &  & $6.13\times10^{34}$ &  & $1.43\times10^{35}$ &  & $2.06\times10^{34}$ &  \\
No-Clump & 90 & $1.66\times10^{33}$ &  & $5.97\times10^{33}$ &  & $6.42\times10^{34}$ &  & $1.54\times10^{35}$ &  & $4.91\times10^{34}$ &  \\
 & 135 & $1.15\times10^{33}$ &  & $3.88\times10^{33}$ &  & $1.92\times10^{34}$ &  & $1.50\times10^{34}$ &  & $1.41\times10^{34}$ &  \\
\hline
 & 45 & $1.27\times10^{33}$ & 1.6 & $4.60\times10^{33}$ & 2.0 & $6.92\times10^{34}$ & 12.9 & $2.38\times10^{35}$ & 66.4 & $3.24\times10^{34}$ & 57.3 \\
Small Clump & 90 & $1.75\times10^{33}$ & 5.4 & $6.10\times10^{33}$ & 2.2 & $5.53\times10^{34}$ & -13.9 & $1.57\times10^{35}$ & 1.9 & $4.66\times10^{34}$ & -5.1 \\
 & 135 & $1.98\times10^{33}$ & 72.2 & $6.73\times10^{33}$ & 73.5 & $3.14\times10^{34}$ & 63.5 & $2.86\times10^{34}$ & 90.7 & $2.32\times10^{34}$ & 64.5 \\
\hline
 & 45 & $2.05\times10^{33}$ & 64.0 & $6.38\times10^{33}$ & 41.5 & $5.49\times10^{34}$ & -10.4 & $1.35\times10^{35}$ & -72.0 & $1.52\times10^{34}$ & -83.5 \\
Big Clump & 90 & $3.25\times10^{33}$ & 90.4 & $1.01\times10^{34}$ & 69.2 & $7.43\times10^{34}$ & 15.7 & $1.71\times10^{35}$ & 9.1 & $3.80\times10^{34}$ & -17.5 \\
 & 135 & $4.07\times10^{33}$ & 253.9 & $1.21\times10^{34}$ & 211.9 & $5.24\times10^{34}$ & 172.9 & $6.05\times10^{34}$ & 303.3 & $5.16\times10^{34}$ & 266.0 \\
\hline
% END RECEIVE ORGTBL emission
\end{tabular}}
\end{table*}

The non-thermal emission was computed for three different stellar wind scenarios: the steady state of the two-wind interaction structure with no clump, the case with a small clump ($\chi = 10$, $R_{\rm c} = 8\times 10^{10}$~cm), and the case with a large clump ($\chi = 10$, $R_{\rm c} = 4\times 10^{11}$~cm). Also, in addition to considering two magnetic field cases ($\chi_B=10^{-3}$ and $0.1$), three representative viewing angles were considered: $\phi=45^\circ$; $90^\circ$; and $135^\circ$; the first angle corresponding for instance to the superior conjunction of the compact object (SUPC) and a system inclination of $45^\circ$, the second one to an intermediate orbital phase, and the third one might represent the inferior conjunction (INFC) for the same inclination; the parameter values are listed in Table~\ref{tab:freepara}.
The simulation times adopted for the emission calculations, of the cases including a clump, were chosen such that the clump was at its closest point from the pulsar; the state of the emitting flow can be considered steady given the short time needed for the particles to leave the grid. 
For each of these scenarios, we obtained the corresponding synchrotron and IC SEDs. These SEDs are presented in Figs.~\ref{steady-3angles}--\ref{lines}, the last one showing the contribution from each streamline.  Figure~\ref{dist} shows the particle energy distribution in the LF for each streamline, and the summation of all of them. Maps were also computed to show how the emission is distributed in the $rz$-plane in the shocked pulsar wind region. These maps are presented in Figs.~\ref{steadymaps} and \ref{bigmaps}. Figure~\ref{special} illustrates the importance of the extended nature of the emitter comparing two SEDs: those obtained for the computed emitter geometry, and those computed assuming that all the emitting cells have the pulsar location (keeping, nonetheless, the particle distributions of the extended emitter). The extended emitter on the scales studied has a minor effect on IC but a major one in gamma-ray absorption for the represented case with $\phi=45^\circ$.

The impact of energy losses on the non-thermal particles can be seen comparing the injected non-thermal luminosities with the energy leaving the computational domain per time unit in the LF. Particles lose $1.6\times 10^{35}$~erg~s$^{-1}$ through radiative losses, or a $\approx 23\%$ of the injected non-thermal luminosity in the no-clump case, whereas in the large-clump case, $1.8\times 10^{35}$~erg~s$^{-1}$ are radiated, or a $\approx 20\%$. The adiabatic losses have more impact, around a $49\%$ of the injected non-thermal luminosity  for both no-clump and big clump cases. The total injected non-thermal luminosities would be $7.1\times 10^{35}$ and $9.1\times 10^{35}$~erg~s$^{-1}$ for the cases without clump and with a large clump, respectively. Doppler boosting shows up comparing the synchrotron+IC total luminosity in the observer frame, for the case without clump, $\phi=90^\circ$, and $\chi_B=10^{-3}$, of $2.75\times 10^{35}$~erg~s$^{-1}$, with the same quantity in the fluid frame, of $1.2\times 10^{35}$~erg~s$^{-1}$, where there is an increase of a factor between 2 and 3 in the emission. 

For informative purposes and completeness, we provide in Table~\ref{tab:bands} the luminosities in the energy bands $1-10$~keV, $10-100$~keV, $0.1-100$~MeV, $0.1-100$~GeV, and $0.1-100$~TeV for the cases without and with clump, focusing on low magnetic field case ($\chi_B=10^{-3}$) that may be considered the more realistic one (see Sect.~\ref{ntemit}). It is clear that in the system considered, which is rather representative \citep{dub13}, the radiative cooling represents about $\sim 1/3$ of the injected non-thermal luminosity, and for $\eta_{\rm NT}=0.1$, about a few \% of $L_{\rm p}$. One can also observe the tendency, suggested in \cite{Bos13}, that in the large-clump case adiabatic losses get enhanced because of the shrinking of the two-wind interaction region. This effect can be seen as a suppression of the lower energy part (where adiabatic losses are important) of the IC spectra in Fig.~\ref{90-lowhigh}. 

\subsection{Applications to \psr}

The binary pulsar system \psr consists of a 47.7~ms radio pulsar \citep{jml92,kdl11} on an eccentric orbit around a luminous O8.5\,Ve star \citep{nrh11}. The pulsar orbits is characterised by the following orbital parameters: eccentricity $e=0.87$, period $P_{{\rm orb}}= 3.5$~yr, and semi-major axis $a_{\rm 2}=6.9\rm \, AU$ \citep[see][and references therein]{nrh11}. The stellar companion in the system is thought to rapidly rotate \citep{nrh11}. The rotation results in a strong surface temperature gradient ($T_{\rm pole}=3.4\times10^4$~K and $T_{\rm eq}=2.75\times10^4$~K), and in the formation of a circumstellar disc. The plane of the orbit and the disc plane are expected to be  misaligned. \citet{nrh11} derived the orbital inclination of $i_{\rm orb}\approx23^\circ$ and adopted that the star rotation axis is inclined in respect to the line of sight by $32^{\circ}$ (i.e., the star is mostly seen from the pole). Accounting for the uncertainty of the star orientation, the IC emission/loss process can be well approximated by a blackbody with temperature $T_*=3\times10^4$~K \citep{kab11}.

  The temperature and luminosity of the UV companion correspond to a late O-star, but because of the fast rotation a Be-star type disc is formed. Thus the stellar wind in \psr should consist of two distinct regions: a dense Keplerian disc, which is crossed by the pulsar approximately $-16$ and $+18$ days to periastron passage, and a fast low density polar wind. The mass-loss rate and density of the discs around Be-stars can vary significantly, but the density should be significantly higher than in the polar wind, and the azimuthal velocity should be Keplerian, i.e., $v_{\rm d}\sim 2\times10^2\rm\,km\,s^{-1}$ (at the distance of $10^{13}\rm \,cm$ from the star). Based on the shape of the X-ray light curve, \citet{ton12} augured that the density of the disc base in \psr might be very high $\rho_0\sim10^{-9}\rm\,g\,cm^{-3}$, but, probably, this conclusion is a subject of large computational uncertainties. The polar wind should be similar to a wind from a O-type star, $v_{\rm w}\sim 2\times10^3\rm\,km\,s^{-1}$. Both these winds might be inhomogeneous, but the disc clumps can be significantly more massive, so in what follows we consider the possible impact of a disc clump on the non-thermal emission in the system.

The system displays variable broadband non-thermal radio, X-ray, and TeV gamma-ray emission close to periastron passage \citep{gtp95,jbw05,cnl06,aaa05,aaa09,utt09,haa13,can14,cn15,ccl15,rbm15}. Although the light curves of the non-thermal emission clearly diverge in different energy bands, the general tendency is similar. Approximately 30 days to periastron passage the flux starts to increase reaching its maximum around 20-10 days before periastron passage (depending on the energy band). When the pulsar approaches periastron, a decreasing tendency in the flux level is apparent. In post-periastron epoch flux increases again, and then gradually decreases. These two-hump structure is less pronounced in the radio band, probably, because of the long cooling time of the radio-emitting electrons, and in TeV gamma rays, possibly because of larger data uncertainties. The X-ray light curve displays quite stable orbit-to-orbit behavior with two clear maxima. The pre- and post-periastron maxima of the X-ray light curve are characterised by similar patterns: a sharp increase and a slower decay, with a larger maximum flux level in the post-periastron epoch \citep{cn15}. The X-ray peaking fluxes are reached close to the epochs, when the pulsed radio emission disappears. The weakening of the pulsed emission is conventionally associated with the pulsar eclipse by the circumstellar disc approximately during the epoch $(-16,+18)$ days to periastron passage.  The location of the light curve maxima suggests that the circumstellar disc may play an important role in the formation of the non-thermal emission. In turn, as indicated by the change of the equivalent width of $\mathrm H\alpha$ \citep{cn15}, the pulsar also affects the disc. This complex interplay makes modelling of the emission in this system to be a very challenging task, thus a consistent multiwavelength interpretation is currently missing \cite[see, however,][]{ton12}.

In addition to TeV gamma rays, bight GeV flares have been detected from the system with \fermi in 2011 January and 2014 May \citep[see][and references therein]{ccl15}. In both epochs the onset of the flares occurred approximately one month after periastron passage. During the flare summit the measured GeV luminosity in 2011 was a factor of $\sim1.5$ higher than, but the flare duration was similar to, that in 2014 \citep[see, e.g.,][]{rbm15}. The typical flare flux is $\sim10^{-6}\rm \,ph\,cm^{-2}s^{-1}$, which means a GeV luminosity $\sim 10\%$ the pulsar spin-down one ($L_{\rm p}\approx 8\times 10^{35}$~erg~s$^{-1}$, \citealt{jml92}) for the distance to the system of $2.3\rm\, pc$ \citep{nrh11}. No detectable change of the TeV emission was observed during the onset of the GeV flare in 2011 \citep{haa13}. Hard X-ray emission detected with \nustar during the GeV flare suggests that the GeV emission might be generated by the same radiation mechanism as the multi keV X-ray emission, namely through the synchrotron channel \citep{tlt15,cn15}.

In what follows we qualitatively consider the possible contribution of the emission generated by the interaction of stellar wind clumps with the pulsar wind to the radiation detected from \psr, for the GeV flare, and the disc crossing in TeV. 

\subsubsection{The GeV flare}

As discussed in \cite{pbp15}, the GeV flare detected from \psr by \fermi could be related to the impact of a dense, big clump of material, probably associated to the decretion disc of the Be star \citep{can14}. Such a GeV flare seems to follow a repetitive pattern \citep[e.g.][]{ccl15,cn15}, which in the scenario just sketched would imply that the disc is similarly affected orbit to orbit, and parts of it are teared apart and directed towards the pulsar. The ram pressure of such a piece of circumstellar disc could have a strong impact on the two-wind interaction structure. As shown by \cite{pbp15}, just a small fraction of the disc mass in the form of a matter clump would be enough to strongly reduce the size of the interaction region. As it was outlined above the flare onset occurs approximately two weeks after the reappearance of the pulsed radio signal, and thus the pulsar has moved relatively far from the disc. We nevertheless assume that a dense cloud
can still collide with the pulsar wind at this epoch. 

The arrival of a large, dense piece of disc could potentially enhance, by compressing the emitting region, the energy density of the local radiation field (e.g. X-rays from the collision region, \citealt{DubCer13}; see also \citealt{kab12} for a related scenario based on IR target photons) to a point when synchrotron self-Compton (SSC) becomes dominant over external IC. This process seems nevertheless unlikely in \psr. Given the pulsar spin-down and the stellar luminosities ($L_*\approx 3\times 10^{38}$~erg~s$^{-1}$,\citealt{nrh11}), even optimistically assuming that most of $L_{\rm p}$ goes to the target photons for SSC, the Thomson approximation for IC, and neglecting IC angular effects, to increase the local photon energy density above the stellar level the two-wind interaction region should get $\lesssim 1/40$ times the size typically considered. This is $\sim 10$\% of the pulsar-star separation distance (\citealt{DubCer13} obtained a similarly small emitting region). Although a detailed account of this possibility is still lacking, such a huge reduction in size requires an increase in stellar material ram pressure by several orders of magnitude and does not seem plausible.

The arrival of a clump may by itself be enough to substantially enhance the IC emission for a combination of Doppler boosting, pulsar wind shock obliqueness, etc. To check this possibility, we computed the emission of the steady and the large clump cases adopting the parameters of \psr at the time of the flare in GeV (not shown here). We took the orbital phase to be around the inferior conjunction of the compact object, an inclination of $23^\circ$, and scaled the hydrodynamic solution \citep{pbp15} to a star-pulsar separation distance in accordance with the source properties. This scaling yields a clump mass $\sim 10^{21}$~g, and a clump destruction time of about a week (the actual flare lasted for about few weeks). For instance, in the low $B$ case, a jump in gamma-ray luminosity by a factor of $\sim 2$ was obtained, not far from the difference between the flare and the periastron \fermi luminosity in \psr \citep[e.g.][]{ccl15}. Unfortunately, the energetics in our case was short by about two orders of magnitude, even when adopting $\chi_{\rm NT}=1$. Therefore, the simulated scenario is in the present form far from being able to explain the GeV flare in \psr. 
%%Thus, we have to conclude in the framework of the considered scenario that the emission enhancement at the orbital phase corresponding to $\sim 30$ days to periastron passage is not caused by the pulsar-disc interaction \cite[see, however,][]{ton12}. If the pulsar interacts with the polar wind during the GeV flare then this phenomena may have a similar origin to multi MeV emission that dominates the SED in \ls, and currently lacks a successful explanation \citep[see][for the most detailed modelling to date]{dlf15}. 

\subsubsection{Disc crossing and TeV emission}

We studied also the interaction of the pulsar wind with the Be star disc assuming that this may be roughly approximated as the encounter between the the former and a large inhomogeneity. From a hydrodynamical point of view, as shown in \cite{onn11}, the disc may be of great dynamical importance for the geometry of the pulsar wind termination shock, and thus in the overall non-thermal emitting region \citep[see][for non-relativistic calculations without particle energy losses]{ton12}. The effects of the disc in the non-thermal emission in \psr had been earlier discussed, for instance in \cite{kha07} (see also \citealt{cnl06}).

We computed the SED for both the steady and the large clump cases when the pulsar is supposed to cross the Be disc after periastron passage. Assuming that the beginning of the disc crossing could be roughly modeled by the encounter with a large clump, we tried to semi-quantitatively reproduce the energetics and increase in flux above 100~GeV observed around those orbital phases. As it was slightly improving results, and given the inclination uncertainty, we adopted $i=30^\circ$ instead of $23^\circ$. As shown in Fig.~\ref{discpas}, with minor changes in the calculation setup and little parameter tuning ($\eta_{\rm NT}=1$; $\chi_B=10^{-3}$)\footnote{Note nevertheless the needed high non-thermal efficiency.}, similar flux levels and evolution (a rise by a factor of several) are obtained, although with a somewhat steeper spectrum than the average one given in \cite{rbm15}.

\begin{figure*}
\centering
\resizebox{0.95\hsize}{!}{\includegraphics{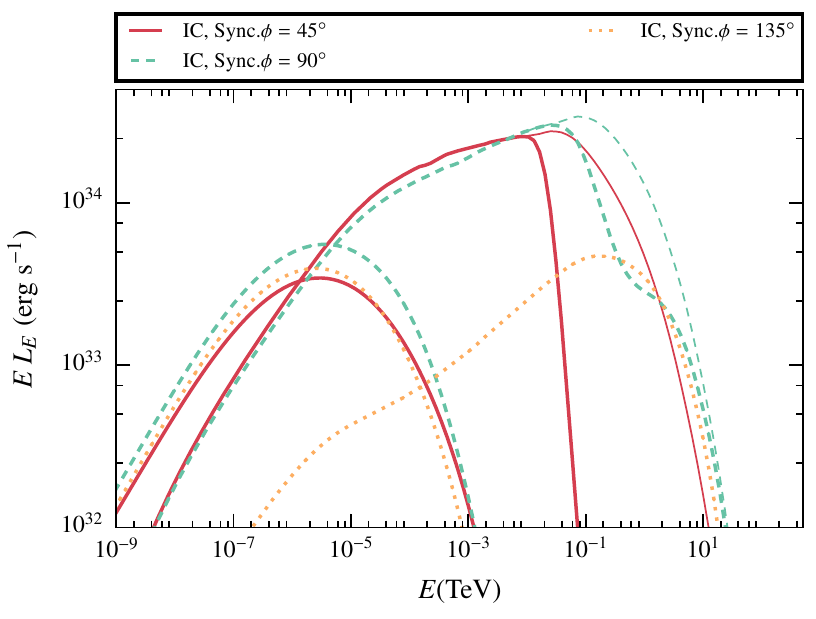}\includegraphics{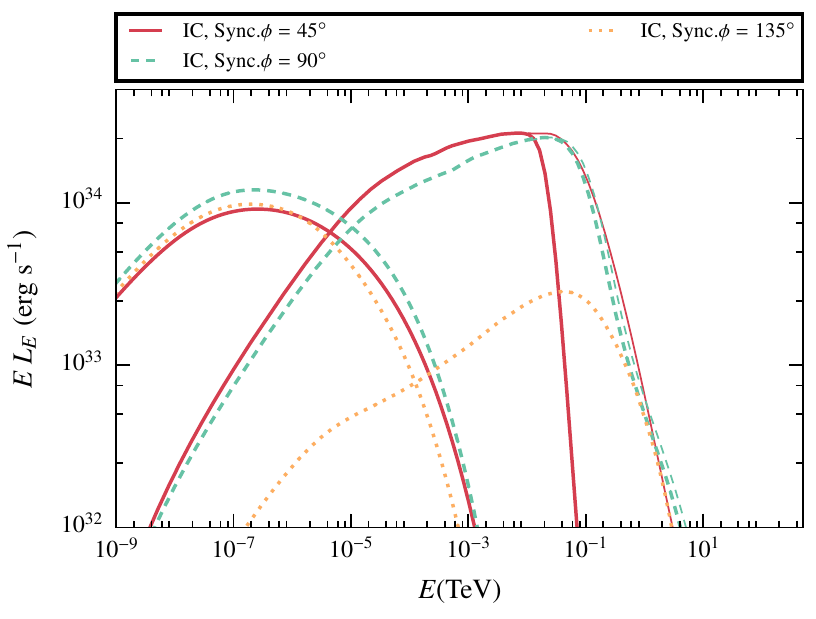}}
\caption{Spectral energy distribution of synchrotron and IC for the no-clump scenario, in the low ($\chi_B = 10^{-3}$) and high ($\chi_B = 0.1$) magnetic field cases (left and right, respectively). Notice that the unabsorbed (thin lines) and the absorbed (thick lines) IC emission are distinguishable only for $\phi=45^\circ$ and 90$^\circ$.}
\label{steady-3angles}
\end{figure*}

\begin{figure*}
\centering
\resizebox{0.95\hsize}{!}{\includegraphics{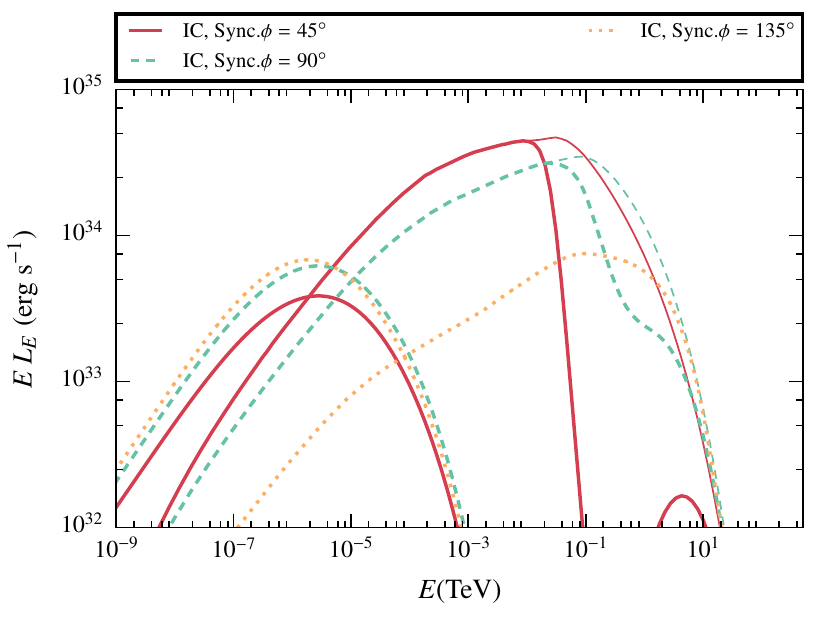}\includegraphics{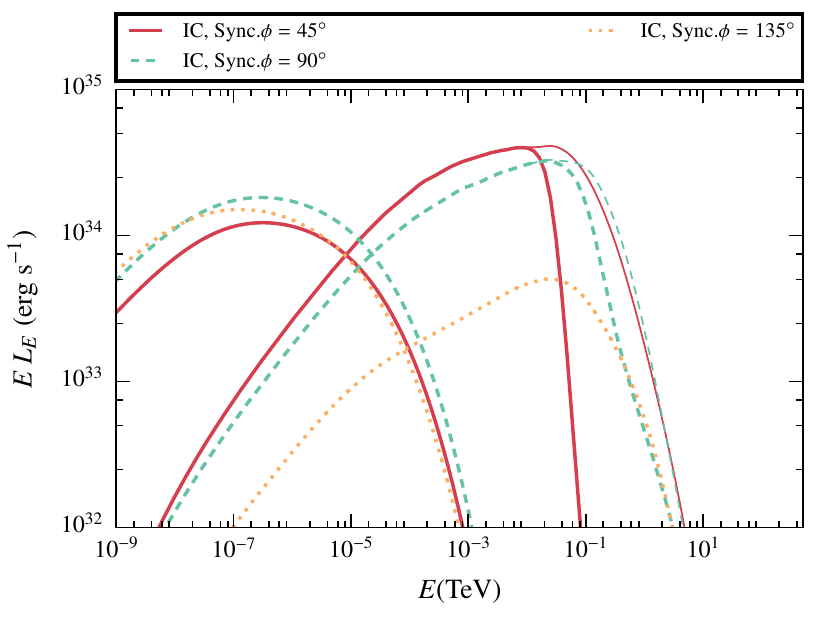}}
\caption{Same as Fig.~\ref{steady-3angles}, but for the small clump ($\chi = 10$, $R_{\rm c} = 8\times 10^{10}$~cm).}
\label{small-3angles}
\end{figure*}

\begin{figure*}
\centering
\resizebox{0.95\hsize}{!}{\includegraphics{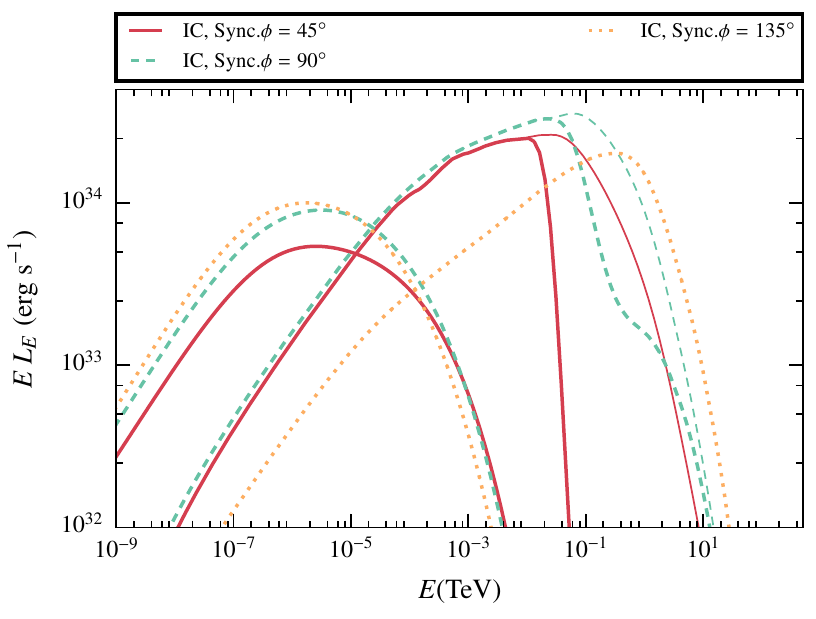}\includegraphics{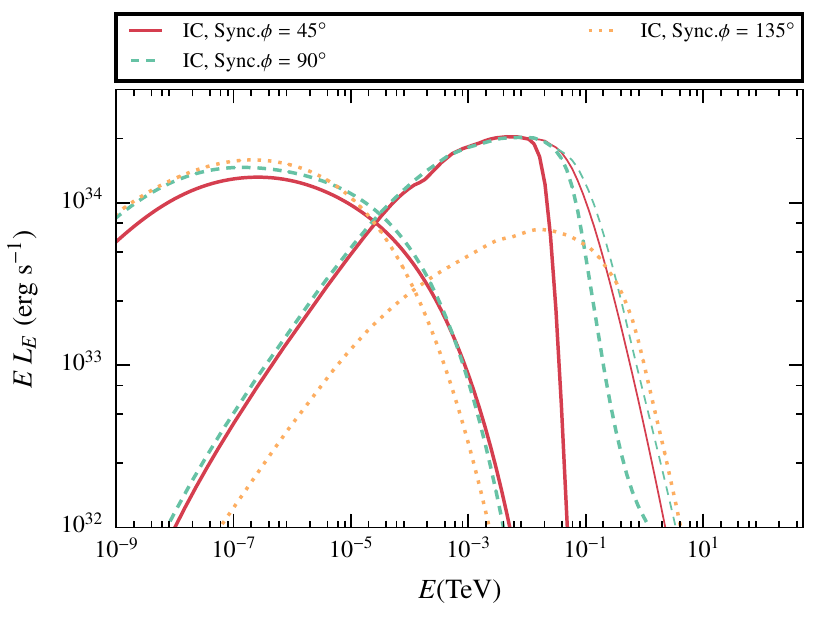}}
\caption{Same as Fig.~\ref{steady-3angles}, but for the large clump ($\chi = 10$, $R_{\rm c} = 4\times 10^{11}$~cm).}
\label{big-3angles}
\end{figure*}

\begin{figure*}
\centering
\resizebox{0.95\hsize}{!}{\includegraphics{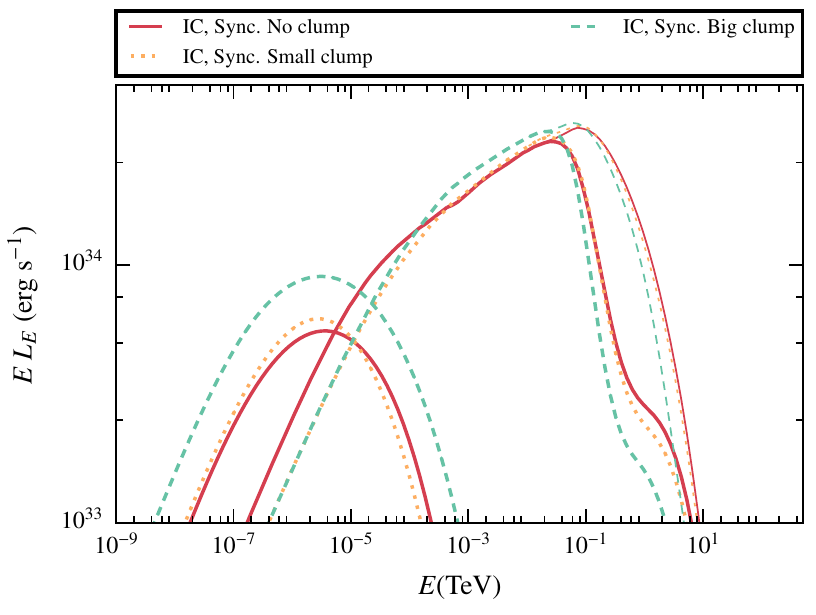}\includegraphics{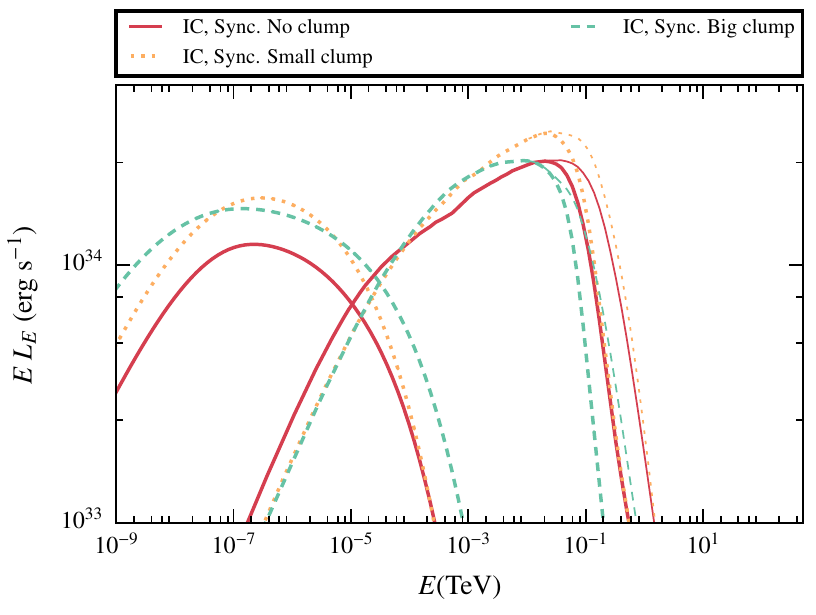}}
\caption{Spectral energy distribution of synchrotron and IC for $\phi=90^\circ$, in the low ($\chi_B = 10^{-3}$) and high ($\chi_B = 0.1$) magnetic field cases (left and right, respectively).}
\label{90-lowhigh}
\end{figure*}

\begin{figure*}
\centering
\resizebox{0.95\hsize}{!}{\includegraphics{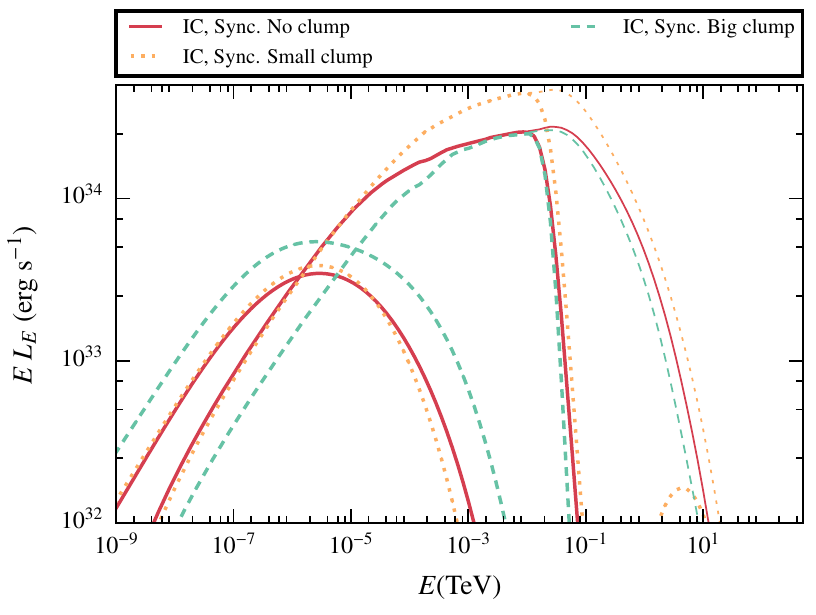}\includegraphics{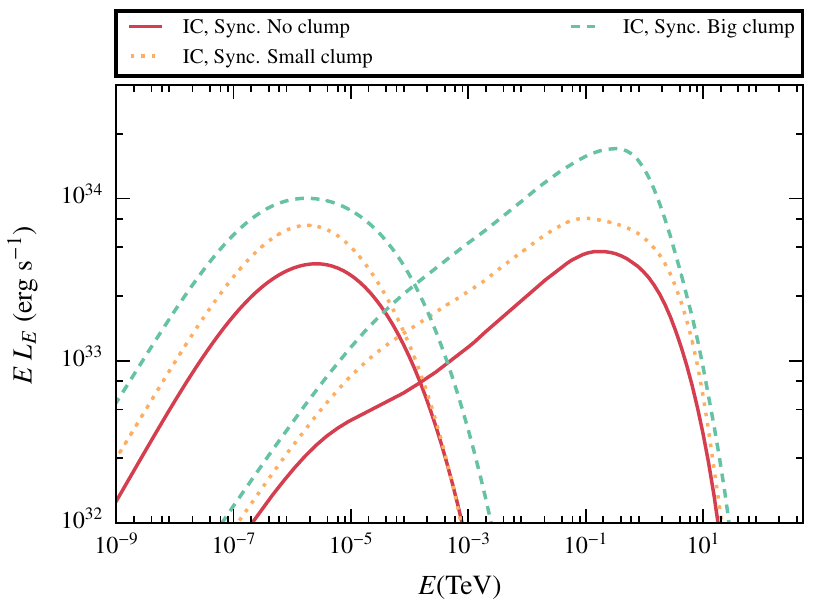}}
\caption{Spectral energy distribution of synchrotron and IC for a fixed magnetic field (low case, $\chi_B = 10^{-3}$) and different $\phi$-values: $45$ and $135^\circ$ (left and right, respectively).}
\label{low-45-135}
\end{figure*}

\begin{figure*}
\centering
\resizebox{0.95\hsize}{!}{\includegraphics{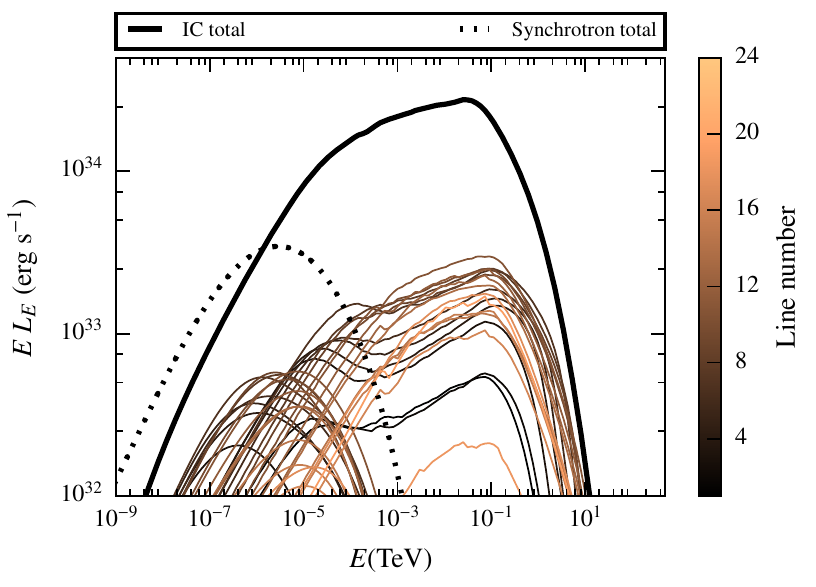}\includegraphics{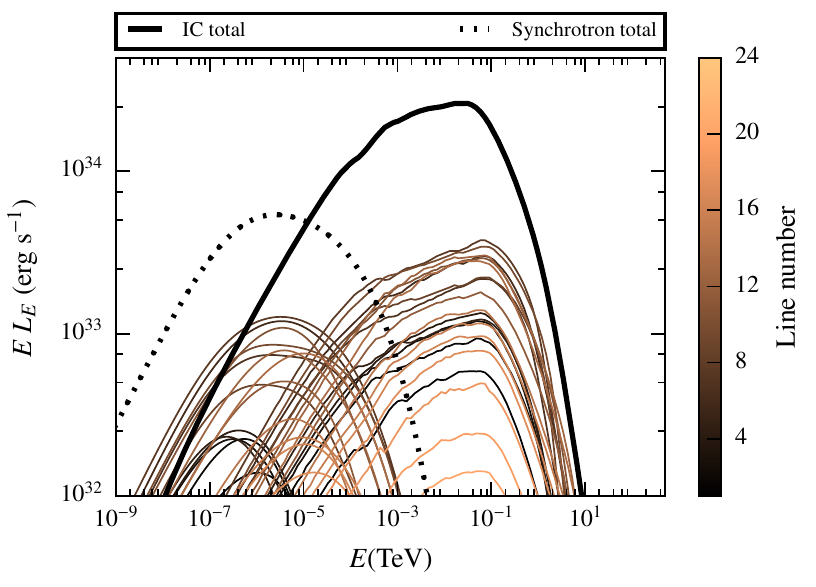}}
\caption{Contribution of each streamline (numbered as in left panel of Fig.~\ref{steady}) to the overall synchrotron (dotted line) and IC (solid line) SED. Both no-clump (left) and large-clump (right) scenarios are computed for a low magnetic field ($\chi_B = 10^{-3}$) and $\phi=90^\circ$.}
\label{lines}
\end{figure*}

\begin{figure}
\centering
\resizebox{0.95\hsize}{!}{\includegraphics{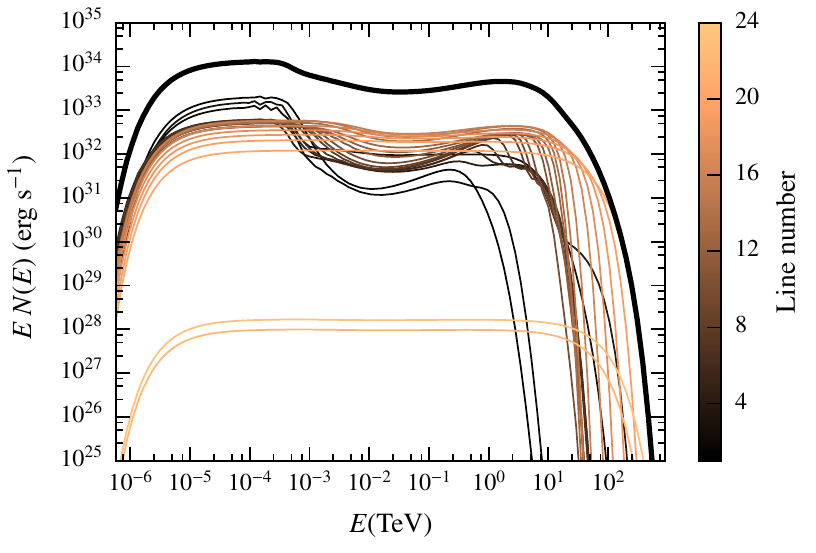}}
\caption{Particle energy distributions for the different streamlines (thin lines) and the sum of all of them (thich black line) in the LF, for the low magnetic field ($\chi_B = 10^{-3}$) no-clump scenario.}
\label{dist}
\end{figure}

\begin{figure*}
\centering
\resizebox{0.95\hsize}{!}{\includegraphics{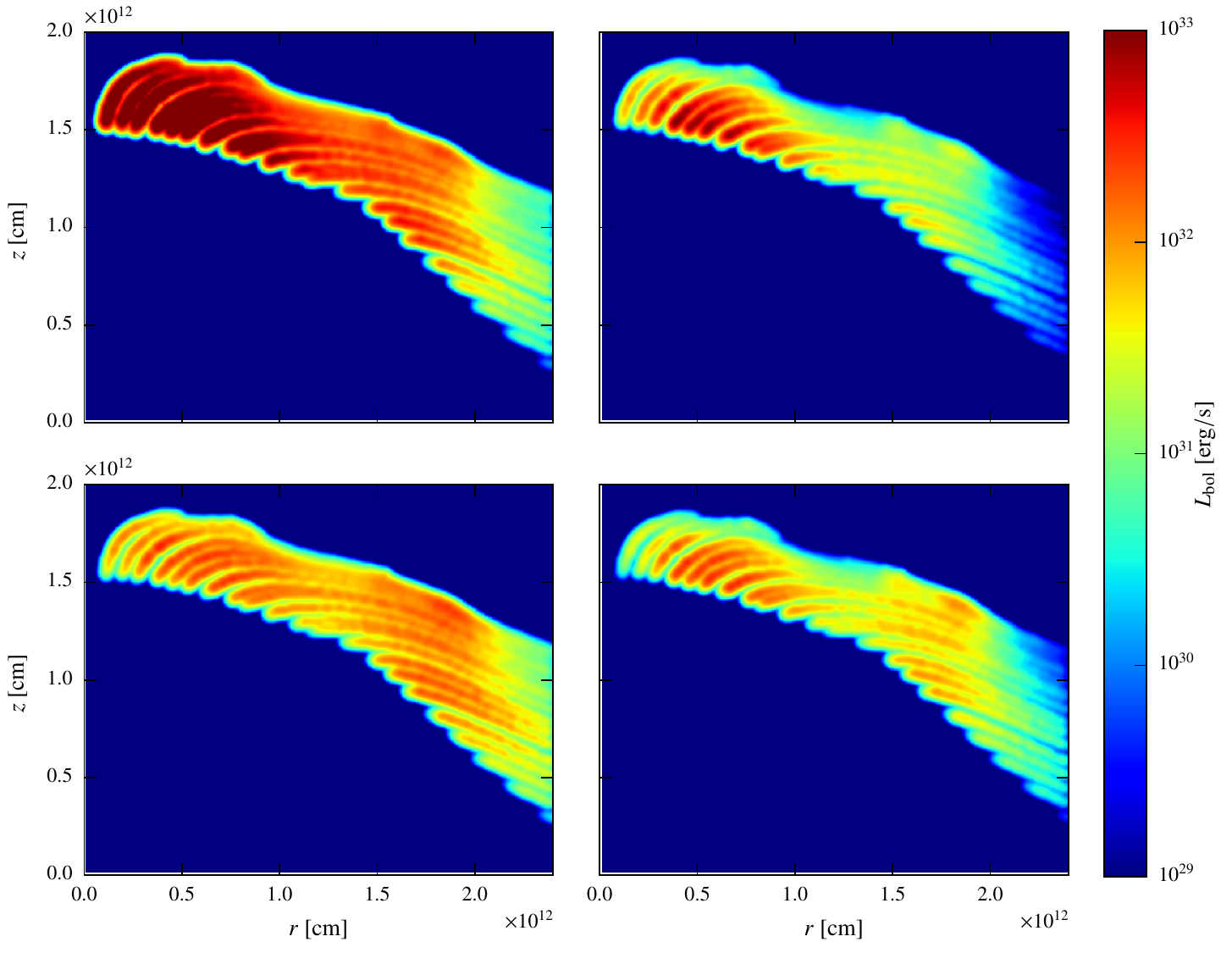}}
\caption{Maps of the bolometric luminosity per cell, for the no-clump scenario, taking a low magnetic field $\chi_B=10^{-3}$. Top panels: map in the $rz$-plane of the distribution of IC (left) and synchrotron (right), for $\phi= 45^\circ$. Bottom panels: same as the top panels but for $\phi= 135^\circ$.}
\label{steadymaps}
\end{figure*}

\begin{figure*}
\centering
\resizebox{0.95\hsize}{!}{\includegraphics{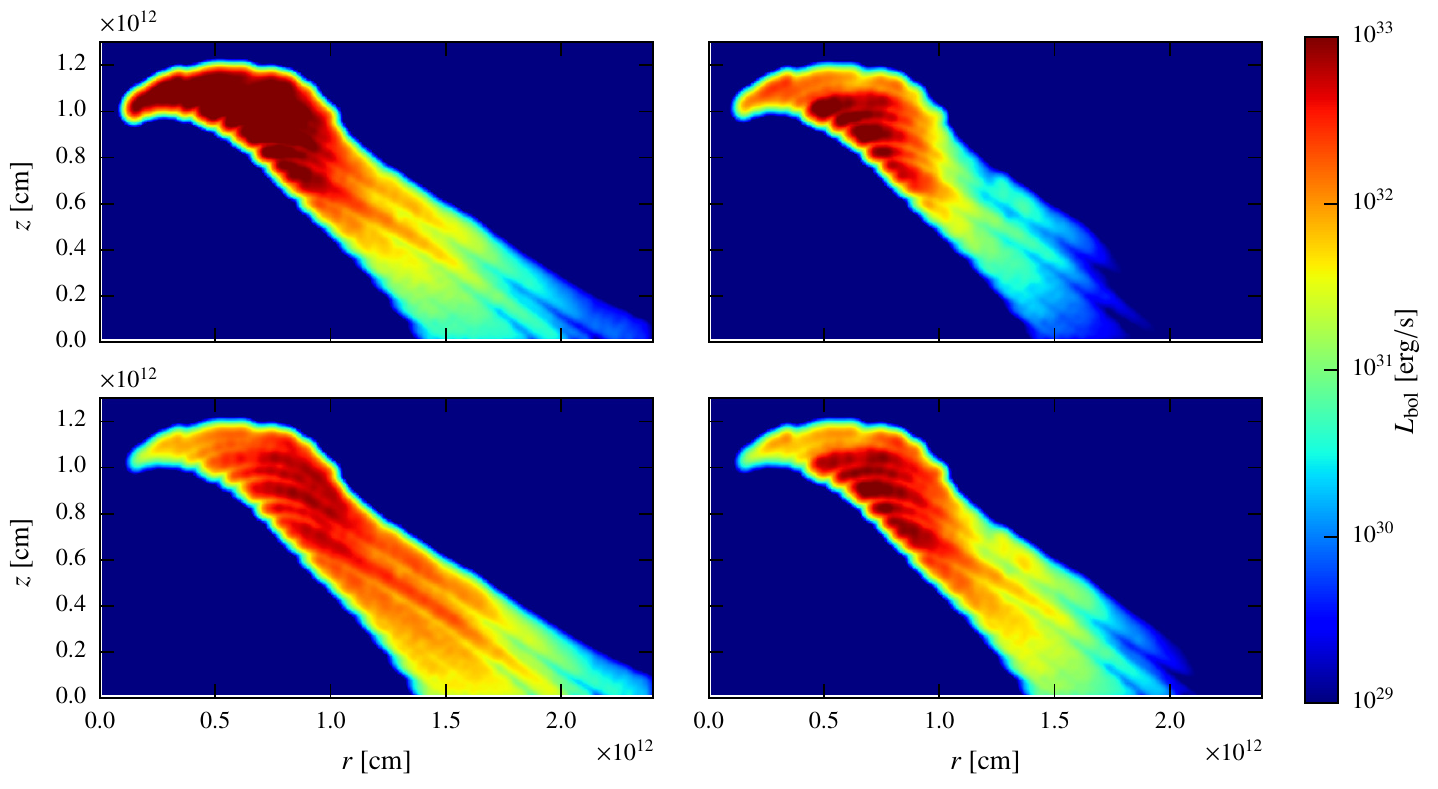}}
\caption{Maps of the bolometric luminosity per cell, for the large clump scenario, taking a low magnetic field $\chi_B=10^{-3}$. Top panels: IC (left) and synchrotron radiation (right) for $\phi= 45^\circ$. Bottom panels: same for $\phi= 135^\circ$.}
\label{bigmaps}
\end{figure*}

%In all the cases, the radiation is partially absorbed by the creation of electron-positron pairs by the interaction with the photon field of the companion star. This absorption is computed taking into account the spacial distribution of the shocked region, as the light travels from the emitter cell to the observer. Is noteworthy that if we considered the emitter to be punctual, the absorption due to pair creation would be different, as shown in Fig.~\ref{special}.

\begin{figure}
\centering
\resizebox{0.95\hsize}{!}{\includegraphics{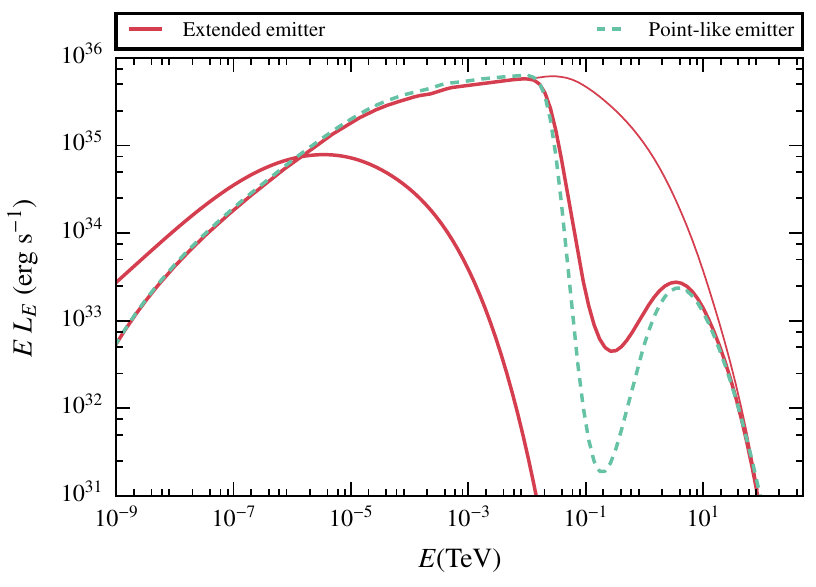}}
\caption{Spectral energy distribution of the synchrotron and IC emission in the no-clump scenario, with $\phi=45^\circ$ and $\chi_B=10^{-3}$. The emission for the computed emitter geometry (solid line) and that obtained assuming that the emitter is point-like (dashed line) are shown. The lower energy component corresponds to synchrotron emission.}
\label{special}
\end{figure}

\begin{figure}
\centering
\resizebox{1.05\hsize}{!}{\includegraphics{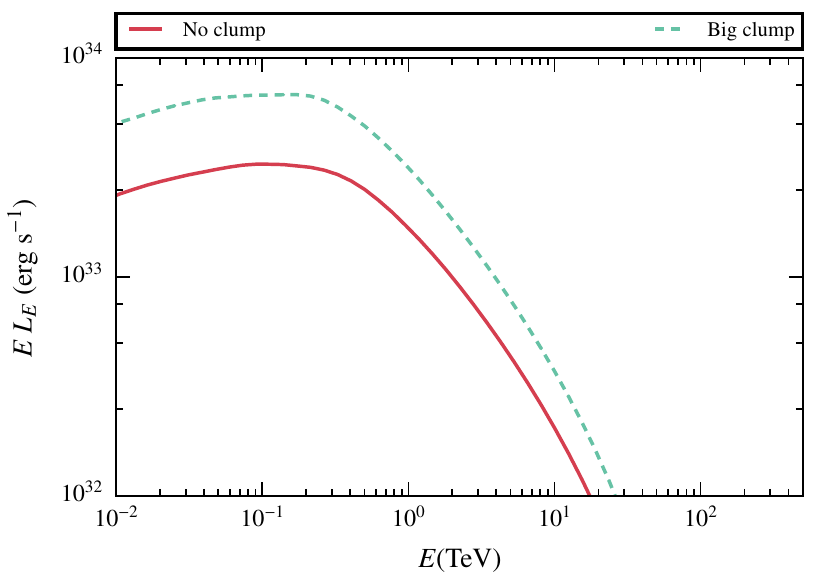}}
\caption{Comparison of the spectral energy distribution for the steady and the large clump case adapting the hydrodynamical results to the case of \psr around 20~days after periastron, roughly at the second
disc passage.}
\label{discpas}
\end{figure}

%%%%%%%%%%%%%%%%%%%%%%%%%
\section{Conclusions}\label{sec:conclusions} 

The results shown in Figs.~\ref{steady-3angles}--\ref{dist} are partially determined by several well known effects: (i) There is a competition between synchrotron and IC losses at particle energies high enough for radiation cooling to dominate over adiabatic losses. Different magnetic-to-target radiation energy density ratios yield different synchrotron-to-IC luminosity ratios. Dominant synchrotron, or IC in Thomson regime (for IC photons $\lesssim 10$~GeV), cooling also lead to a spectral softening, whereas dominant IC in Klein Nishina (for IC photons $\gtrsim 10$~GeV) leads to a spectral hardening \citep[e.g.][]{KhaAha05}. (ii) Another important factor is IC scattering on an anisotropic target photon field, which softens and boosts the gamma-ray emission for close-to-head on collisions (around SUPC), and hardens and reduces the gamma-ray emission for small scattering angles (around INFC) \citep[e.g.][]{kab08,dch08}. (iii) At the highest energies, there is absorption through pair creation in the anisotropic photon field of the star. This process has its minimum threshold energy of the absorbed gamma rays ($\sim m_e^2 c^4/2.7\,kT_\star\approx 30$~GeV), and the strongest attenuation, for close-to-head on collisions (around SUPC), whereas it presents a high absorption threshold, and weak attenuation, for small photon-photon interaction angles (around INFC) \citep[e.g.][]{BotDer05,dub06,kab08}. For the adopted stellar luminosity and orbital separation distance, gamma-ray absorption around 1~TeV is very strong for $\phi<90^\circ$. This suggests that the bright TeV emission detected from several close binary systems, e.g. \ls\, \citep{aab06}, cannot be entirely generated in the inner part of the colliding wind structure, and some additional production sites should be considered \citep{zba13}.

The importance of synchrotron losses was discussed in \cite{Bos13} in the context of a clump large enough to significantly reduce the size of the two-wind interaction region, which is expected to enhance the magnetic field in the shocked pulsar wind. This effect is seen in our results (see Table~\ref{tab:bands}) when comparing the synchrotron and IC luminosities in different energy bands, and under the absence or presence of clumps. Adiabatic losses are also increased by the reduction of the two-wind interaction region, as noted in Sect.~\ref{resu}.

Doppler boosting is a very important factor characterizing the observer synchrotron and IC luminosities (by $\delta^4$), and to a lesser extent spectra through photon energy boosting (by $\delta$), in the scenario studied here. Doppler boosting was already discussed in the past \citep[e.g.][]{kab08,dch10,kch12,kba14,br15,dlf15} in the framework of a homogeneous stellar wind, in which context we obtain similar results, as Doppler boosting induces luminosity variations of up to a factor of a few comparing different viewing angles in certain bands (see Table~\ref{tab:bands}, no-clump scenario). Radiation enhancement in the cases computed here is well illustrated by the maps presented in the right panels of Figs~\ref{steady}, \ref{Chi10Rc1}, and \ref{Chi10Rc5}. For complementarity, the clump absence/presence impact is also shown in the left panel of Figs.~\ref{steady}, \ref{Chi10Rc1}, and \ref{Chi10Rc5} for the general direction of the streamlines, and in the central panel of the same figures. for the shocked pulsar wind speed distribution. The effects on radiation are seen in the SEDs and maps shown in \ref{steadymaps} and \ref{bigmaps} (and again in Table~\ref{tab:bands}, comparing the same energy bands and $\phi$-values for the three cases). 

If the stellar wind is homogeneous, the two-wind interaction structure is already prone to suffer hydrodynamical instabilities \citep[e.g.][]{bbk12,lfd13,pbp15,bbp15}, i.e. the tinniest irregularities in the wind act as inhomogeneity seeds. The corresponding dispersion in the velocity field coupled with Doppler boosting can thus introduce a chaotic variability component to the emission even in the absence of  clumps. Therefore, even rather small clumps enhance instability formation and growth, making the emission variability further complex. 

It is interesting to compare the cases without and with a small clump in Table~\ref{tab:bands} for the same viewing angle. In these two cases, the overall two-wind interaction structure does not change significantly, and the flux difference is determined by the perturbations in the flow velocity. This already leads to luminosity variations of almost a factor of $\sim 2$ in certain bands. Large clumps, besides adding perturbations to the shocked flow structure, also change the overall distribution of the streamline directions as the shock approaches the pulsar significantly. This induces a characteristic pattern, with its evolution determined by the shocked clump dynamics. Note that in the more realistic case of using a 3D simulation to compute the emission, the lack of symmetry in the azimuthal direction with respect to the pulsar-star axis would likely lead to an even more chaotic velocity field, and thus to stronger differences between clump scenarios.

It is worth noting that the shock becomes more perpendicular when approaching the pulsar in the large-clump case. Therefore, the transfer of energy to non-thermal particles is increased, and thus the clump presence enhances particle acceleration at the pulsar wind shock, as shown by the injected luminosity into non-thermal particles, which in the large-clump case is a $\sim 50$\% larger than without clump (see Sect.~\ref{resu}). This effect may be however difficult to disentangle from Doppler boosting in the radiation results. Also, if magnetic field reconnection plays a sensible role in particle acceleration, chaotic perturbations of the magnetic field structure induced by clumps may have a strong impact on the particle acceleration \citep{SirSpi11}.

 From a direct application of our radiation calculations using hydrodynamical information to the post-periastron GeV flare, and the post-periastron disc crossing in TeV, in \psr, the former is hard to reproduce in our simplified scenario, whereas the second semi-quantitatively agrees with our results.

\section{Final remarks}\label{sec:discussion} 

This work provides with some illustrative examples of how different types of clumps can affect the emitting region and the radiation itself in the case of an inhomogeneous stellar wind interacting with a pulsar wind. The studied effects (clump presence, velocity field dispersion, Doppler boosting for a given observer, instability development, magnetic field increase, and closing of the pulsar wind shock) acting together can either cancel out to some extent, or sum up rather unpredictably. There is also a diversity of timescales, as instability growth, region shrinking, and magnetic field growth depend on the clump evolution timescale, which itself depends to first order on the clump size and density. On the other hand, the shocked pulsar wind flow can change direction much faster, and the rapidly changing, nonuniform, beaming of radiation in the emitting region is an important factor shaping variability. One can thus conclude that flares could occasionally be seen in some or all bands of the spectrum, with their duration determined by the dominant variability origin, whereas in general emission may vary more smoothly. There are periodic emission features in the systems studied that originate in repetitive physical phenomena (e.g. orbit-related IC, orbit-related Doppler boosting, pair creation angular effects, changes in radiative and non-radiative cooling along the orbit, etc.), but non-periodic variability originates from a combination of different, equally important, factors, and they can be hard to disentangle. 
A study of the X-ray light curve can be informative of the different processes shaping the non-thermal emission.
We note that \cite{ktu09} found that the X-ray light curve of the pulsar binary candidate {\ls} in the years 1999--2007 was rather stable, with even fine structures such as spikes and dips similar from one orbit to another. This non chaotic behavior, if confirmed, would not be explained by the processes discussed in this work.

A quantitative assessment of the importance of the different factors in the clump wind scenario is a next step to be carried out. The reason is that the shocked pulsar wind accelerates as it propagates \citep[e.g.][]{bkk08}, and it does so in parallel with instability growth. Therefore, a quantitative prediction of the impact of instability growth on the emission, induced either by small perturbations or large clumps, requires a larger computational grid to properly capture all these processes. Multiple clump interactions should be also simulated. In this regard, a stellar wind with a distribution of clump properties \citep[e.g.][]{mof08} is likely to be an additional variability source affecting the non-linear hydrodynamical processes occurring in the two-wind interaction structure. Finally, the dynamical role of the magnetic field cannot be forgotten if the pulsar wind is weakly magnetized, as the flow magnetization may grow in specific regions of the shocked pulsar wind \citep[see, e.g., fig.~6 in][]{bkk12}\footnote{Although the general impact for a moderate magnetization value was found to be negligible in that work.} enough to moderate the development of hydrodynamical instabilities, or induce anisotropy and thus further complexity to their development.  

If non-thermal radiation losses were to be accounted (say $\eta_{\rm NT}\lesssim 1$) when modeling the properties of the shocked pulsar wind, full radiation-(magneto)hydrodynamic simulations should be carried out. We must point out though that, in addition to this effect, one has also to account for the fact that IC emission on stellar photons is already anisotropic in the FF. This anisotropy in the FF implies that there will be momentum lost in the direction of the star, a form of Compton rocket \citep[e.g.][]{ode81}. Thus, if IC radiation is important for the flow internal (non-thermal) energy losses, its dynamical impact will be also important for the emitting flow through the loss of momentum in specific directions. Synchrotron emission may be also anisotropic if the magnetic field were ordered but, unless the field presented a strong gradient, emission by particles moving in opposite directions along the field lines would effectively cancel the momentum loss out.

%\section{Summary}\label{sec:sum}

%%%%%%%%%%%%%%%%%%%%%%%%%
\begin{acknowledgements}
We thank the anonymous referee for his/her constructive and useful comments.
We acknowledge support by the Spanish Ministerio de Econom\'ia y Competitividad (MINECO) under grants
AYA2013-47447-C3-1-P, and MDM-2014-0369 of ICCUB (Unidad de Excelencia 'Mar\'ia de Maeztu') and the Catalan DEC grant 2014 SGR 86.      
% J.M., Paredes
%We also acknowledge support by the ``Generalitat Valenciana'' grant ``PROMETEO-2009-103''.
This research has been supported by the Marie Curie Career Integration Grant 321520.
V.B-R. also acknowledges financial support from MINECO and European Social Funds through a Ram\'on y Cajal fellowship.
X.P.-F. also acknowledges financial support from Universitat de Barcelona and Generalitat de Catalunya under grants APIF and FI (2015FI\_B1 00153), respectively. D.K. acknowledges support by the Russian Science Foundation under grant 16-12-10443.
M.P. is a member of the working team of projects AYA2013-40979-P and AYA2013-48226-C3-2-P, funded by MINECO.
\end{acknowledgements}

\bibliographystyle{aa}
\bibliography{ALLreferences}

\begin{thebibliography}{80}
\expandafter\ifx\csname natexlab\endcsname\relax\def\natexlab#1{#1}\fi

\bibitem[{{Aharonian} {et~al.}(2009){Aharonian}, {Akhperjanian}, {Anton},
  {Barres de Almeida}, {Bazer-Bachi}, {Becherini}, {Behera}, {Bernl{\"o}hr},
  {Bochow}, {Boisson}, {Bolmont}, {Borrel}, {Brucker}, {Brun}, {Brun},
  {B{\"u}hler}, {Bulik}, {B{\"u}sching}, {Boutelier}, {Chadwick},
  {Charbonnier}, {Chaves}, {Cheesebrough}, {Chounet}, {Clapson}, {Coignet},
  {Dalton}, {Daniel}, {Davids}, {Degrange}, {Deil}, {Dickinson},
  {Djannati-Ata{\"i}}, {Domainko}, {O'C.~Drury}, {Dubois}, {Dubus}, {Dyks},
  {Dyrda}, {Egberts}, {Emmanoulopoulos}, {Espigat}, {Farnier}, {Feinstein},
  {Fiasson}, {F{\"o}rster}, {Fontaine}, {F{\"u}{\ss}ling}, {Gabici}, {Gallant},
  {G{\'e}rard}, {Gerbig}, {Giebels}, {Glicenstein}, {Gl{\"u}ck}, {Goret},
  {G{\"o}ring}, {Hauser}, {Hauser}, {Heinz}, {Heinzelmann}, {Henri}, {Hermann},
  {Hinton}, {Hoffmann}, {Hofmann}, {Holleran}, {Hoppe}, {Horns},
  {Jacholkowska}, {de Jager}, {Jahn}, {Jung}, {Katarzy{\'n}ski}, {Katz},
  {Kaufmann}, {Kerschhaggl}, {Khangulyan}, {Kh{\'e}lifi}, {Keogh}, {Klochkov},
  {Klu{\'z}niak}, {Kneiske}, {Komin}, {Kosack}, {Kossakowski}, {Lamanna},
  {Lenain}, {Lohse}, {Marandon}, {Martineau-Huynh}, {Marcowith}, {Masbou},
  {Maurin}, {McComb}, {Medina}, {Moderski}, {Moulin}, {Naumann-Godo}, {de
  Naurois}, {Nedbal}, {Nekrassov}, {Nicholas}, {Niemiec}, {Nolan}, {Ohm},
  {Olive}, {de O{\~n}a Wilhelmi}, {Orford}, {Ostrowski}, {Panter}, {Paz
  Arribas}, {Pedaletti}, {Pelletier}, {Petrucci}, {Pita}, {P{\"u}hlhofer},
  {Punch}, {Quirrenbach}, {Raubenheimer}, {Raue}, {Rayner}, {Renaud}, {Rieger},
  {Ripken}, {Rob}, {Rosier-Lees}, {Rowell}, {Rudak}, {Rulten}, {Ruppel},
  {Sahakian}, {Santangelo}, {Schlickeiser}, {Sch{\"o}ck}, {Schwanke},
  {Schwarzburg}, {Schwemmer}, {Shalchi}, {Sikora}, {Skilton}, {Sol},
  {Spangler}, {Stawarz}, {Steenkamp}, {Stegmann}, {Stinzing}, {Superina},
  {Szostek}, {Tam}, {Tavernet}, {Terrier}, {Tibolla}, {Tluczykont}, {van
  Eldik}, {Vasileiadis}, {Venter}, {Venter}, {Vialle}, {Vincent}, {Vivier},
  {V{\"o}lk}, {Volpe}, {Wagner}, {Ward}, {Zdziarski}, \& {Zech}}]{aaa09}
{Aharonian}, F., {Akhperjanian}, A.~G., {Anton}, G., {et~al.} 2009, \aap, 507,
  389

\bibitem[{{Aharonian} {et~al.}(2005){Aharonian}, {Akhperjanian}, {Aye},
  {Bazer-Bachi}, {Beilicke}, {Benbow}, {Berge}, {Berghaus}, {Bernl{\"o}hr},
  {Boisson}, {Bolz}, {Braun}, {Breitling}, {Brown}, {Bussons Gordo},
  {Chadwick}, {Chounet}, {Cornils}, {Costamante}, {Degrange},
  {Djannati-Ata{\"i}}, {O'C.~Drury}, {Dubus}, {Emmanoulopoulos}, {Espigat},
  {Feinstein}, {Fleury}, {Fontaine}, {Fuchs}, {Funk}, {Gallant}, {Giebels},
  {Gillessen}, {Glicenstein}, {Goret}, {Hadjichristidis}, {Hauser},
  {Heinzelmann}, {Henri}, {Hermann}, {Hinton}, {Hofmann}, {Holleran}, {Horns},
  {de Jager}, {Johnston}, {Kh{\'e}lifi}, {Kirk}, {Komin}, {Konopelko},
  {Latham}, {Le Gallou}, {Lemi{\`e}re}, {Lemoine-Goumard}, {Leroy},
  {Martineau-Huynh}, {Lohse}, {Marcowith}, {Masterson}, {McComb}, {de Naurois},
  {Nolan}, {Noutsos}, {Orford}, {Osborne}, {Ouchrif}, {Panter}, {Pelletier},
  {Pita}, {P{\"u}hlhofer}, {Punch}, {Raubenheimer}, {Raue}, {Raux}, {Rayner},
  {Redondo}, {Reimer}, {Reimer}, {Ripken}, {Rob}, {Rolland}, {Rowell},
  {Sahakian}, {Saug{\'e}}, {Schlenker}, {Schlickeiser}, {Schuster}, {Schwanke},
  {Siewert}, {Skj{\ae}raasen}, {Sol}, {Steenkamp}, {Stegmann}, {Tavernet},
  {Terrier}, {Th{\'e}oret}, {Tluczykont}, {Vasileiadis}, {Venter}, {Vincent},
  {V{\"o}lk}, \& {Wagner}}]{aaa05}
{Aharonian}, F., {Akhperjanian}, A.~G., {Aye}, K.-M., {et~al.} 2005, \aap, 442,
  1

\bibitem[{{Aharonian} {et~al.}(2006{\natexlab{a}}){Aharonian}, {Akhperjanian},
  {Bazer-Bachi}, {Beilicke}, {Benbow}, {Berge}, {Bernl{\"o}hr}, {Boisson},
  {Bolz}, {Borrel}, {Braun}, {Brown}, {B{\"u}hler}, {B{\"u}sching}, {Carrigan},
  {Chadwick}, {Chounet}, {Cornils}, {Costamante}, {Degrange}, {Dickinson},
  {Djannati-Ata{\"i}}, {O'C.~Drury}, {Dubus}, {Egberts}, {Emmanoulopoulos},
  {Espigat}, {Feinstein}, {Ferrero}, {Fiasson}, {Fontaine}, {Funk}, {Funk},
  {F{\"u}{\ss}ling}, {Gallant}, {Giebels}, {Glicenstein}, {Goret},
  {Hadjichristidis}, {Hauser}, {Hauser}, {Heinzelmann}, {Henri}, {Hermann},
  {Hinton}, {Hoffmann}, {Hofmann}, {Holleran}, {Horns}, {Jacholkowska}, {de
  Jager}, {Kendziorra}, {Kh{\'e}lifi}, {Komin}, {Konopelko}, {Kosack},
  {Latham}, {Le Gallou}, {Lemi{\`e}re}, {Lemoine-Goumard}, {Lohse}, {Martin},
  {Martineau-Huynh}, {Marcowith}, {Masterson}, {Maurin}, {McComb}, {Moulin},
  {de Naurois}, {Nedbal}, {Nolan}, {Noutsos}, {Orford}, {Osborne}, {Ouchrif},
  {Panter}, {Pelletier}, {Pita}, {P{\"u}hlhofer}, {Punch}, {Raubenheimer},
  {Raue}, {Rayner}, {Reimer}, {Reimer}, {Ripken}, {Rob}, {Rolland}, {Rowell},
  {Sahakian}, {Santangelo}, {Saug{\'e}}, {Schlenker}, {Schlickeiser},
  {Schr{\"o}der}, {Schwanke}, {Schwarzburg}, {Shalchi}, {Sol}, {Spangler},
  {Spanier}, {Steenkamp}, {Stegmann}, {Superina}, {Tavernet}, {Terrier},
  {Tluczykont}, {van Eldik}, {Vasileiadis}, {Venter}, {Vincent}, {V{\"o}lk},
  {Wagner}, \& {Ward}}]{aab06}
{Aharonian}, F., {Akhperjanian}, A.~G., {Bazer-Bachi}, A.~R., {et~al.}
  2006{\natexlab{a}}, \aap, 460, 743

\bibitem[{{Aharonian} {et~al.}(2006{\natexlab{b}}){Aharonian}, {Anchordoqui},
  {Khangulyan}, \& {Montaruli}}]{aak06}
{Aharonian}, F., {Anchordoqui}, L., {Khangulyan}, D., \& {Montaruli}, T.
  2006{\natexlab{b}}, Journal of Physics Conference Series, 39, 408

\bibitem[{{Aharonian} \& {Atoyan}(1981)}]{AhaAto81}
{Aharonian}, F.~A. \& {Atoyan}, A.~M. 1981, \apss, 79, 321

\bibitem[{{Aharonian} \& {Bogovalov}(2003)}]{AhaBog03}
{Aharonian}, F.~A. \& {Bogovalov}, S.~V. 2003, \na, 8, 85

\bibitem[{{Aharonian} {et~al.}(2012){Aharonian}, {Bogovalov}, \&
  {Khangulyan}}]{abk12}
{Aharonian}, F.~A., {Bogovalov}, S.~V., \& {Khangulyan}, D. 2012, \nat, 482,
  507

\bibitem[{{Bogovalov} {et~al.}(2012){Bogovalov}, {Khangulyan}, {Koldoba},
  {Ustyugova}, \& {Aharonian}}]{bkk12}
{Bogovalov}, S.~V., {Khangulyan}, D., {Koldoba}, A.~V., {Ustyugova}, G.~V., \&
  {Aharonian}, F.~A. 2012, \mnras, 419, 3426

\bibitem[{{Bogovalov} {et~al.}(2008){Bogovalov}, {Khangulyan}, {Koldoba},
  {Ustyugova}, \& {Aharonian}}]{bkk08}
{Bogovalov}, S.~V., {Khangulyan}, D.~V., {Koldoba}, A.~V., {Ustyugova}, G.~V.,
  \& {Aharonian}, F.~A. 2008, \mnras, 387, 63

\bibitem[{{Bosch-Ramon}(2011)}]{bos11}
{Bosch-Ramon}, V. 2011, ArXiv e-prints

\bibitem[{{Bosch-Ramon}(2013)}]{Bos13}
{Bosch-Ramon}, V. 2013, \aap, 560, A32

\bibitem[{{Bosch-Ramon} {et~al.}(2012){Bosch-Ramon}, {Barkov}, {Khangulyan}, \&
  {Perucho}}]{bbk12}
{Bosch-Ramon}, V., {Barkov}, M.~V., {Khangulyan}, D., \& {Perucho}, M. 2012,
  \aap, 544, A59

\bibitem[{{Bosch-Ramon} {et~al.}(2015){Bosch-Ramon}, {Barkov}, \&
  {Perucho}}]{bbp15}
{Bosch-Ramon}, V., {Barkov}, M.~V., \& {Perucho}, M. 2015, \aap, 577, A89

\bibitem[{{Bosch-Ramon} \& {Khangulyan}(2009)}]{BosKha09}
{Bosch-Ramon}, V. \& {Khangulyan}, D. 2009, International Journal of Modern
  Physics D, 18, 347

\bibitem[{{Bosch-Ramon} \& {Khangulyan}(2011)}]{BosKha11}
{Bosch-Ramon}, V. \& {Khangulyan}, D. 2011, \pasj, 63, 1023

\bibitem[{{Bosch-Ramon} {et~al.}(2008){Bosch-Ramon}, {Khangulyan}, \&
  {Aharonian}}]{bka08}
{Bosch-Ramon}, V., {Khangulyan}, D., \& {Aharonian}, F.~A. 2008, \aap, 482, 397

\bibitem[{{B{\"o}ttcher} \& {Dermer}(2005)}]{BotDer05}
{B{\"o}ttcher}, M. \& {Dermer}, C.~D. 2005, \apjl, 634, L81

\bibitem[{{Bucciantini} {et~al.}(2005){Bucciantini}, {Amato}, \& {Del
  Zanna}}]{bad05}
{Bucciantini}, N., {Amato}, E., \& {Del Zanna}, L. 2005, \aap, 434, 189

\bibitem[{{Caliandro} {et~al.}(2015){Caliandro}, {Cheung}, {Li}, {Scargle},
  {Torres}, {Wood}, \& {Chernyakova}}]{ccl15}
{Caliandro}, G.~A., {Cheung}, C.~C., {Li}, J., {et~al.} 2015, \apj, 811, 68

\bibitem[{{Cerutti} {et~al.}(2010){Cerutti}, {Malzac}, {Dubus}, \&
  {Henri}}]{cmd10}
{Cerutti}, B., {Malzac}, J., {Dubus}, G., \& {Henri}, G. 2010, \aap, 519, A81

\bibitem[{{Chernyakova} {et~al.}(2014){Chernyakova}, {Abdo}, {Neronov},
  {McSwain}, {Mold{\'o}n}, {Rib{\'o}}, {Paredes}, {Sushch}, {de Naurois},
  {Schwanke}, {Uchiyama}, {Wood}, {Johnston}, {Chaty}, {Coleiro}, {Malyshev},
  \& {Babyk}}]{can14}
{Chernyakova}, M., {Abdo}, A.~A., {Neronov}, A., {et~al.} 2014, \mnras, 439,
  432

\bibitem[{{Chernyakova} {et~al.}(2006){Chernyakova}, {Neronov}, {Lutovinov},
  {Rodriguez}, \& {Johnston}}]{cnl06}
{Chernyakova}, M., {Neronov}, A., {Lutovinov}, A., {Rodriguez}, J., \&
  {Johnston}, S. 2006, \mnras, 367, 1201

\bibitem[{{Chernyakova} {et~al.}(2015){Chernyakova}, {Neronov}, {van Soelen},
  {Callanan}, {O'Shaughnessy}, {Babyk}, {Tsygankov}, {Vovk}, {Krivonos},
  {Tomsick}, {Malyshev}, {Li}, {Wood}, {Torres}, {Zhang}, {Kretschmar},
  {McSwain}, {Buckley}, \& {Koen}}]{cn15}
{Chernyakova}, M., {Neronov}, A., {van Soelen}, B., {et~al.} 2015, \mnras, 454,
  1358

\bibitem[{{Colella} \& {Woodward}(1984)}]{ColWoo84}
{Colella}, P. \& {Woodward}, P.~R. 1984, Journal of Computational Physics, 54,
  174

\bibitem[{{de la Cita} {et~al.}(2016){de la Cita}, {Bosch-Ramon},
  {Paredes-Fortuny}, {Khangulyan}, \& {Perucho}}]{bp16}
{de la Cita}, V.~M., {Bosch-Ramon}, V., {Paredes-Fortuny}, X., {Khangulyan},
  D., \& {Perucho}, M. 2016, \aap, 591, A15

\bibitem[{{del Palacio} {et~al.}(2015){del Palacio}, {Bosch-Ramon}, \&
  {Romero}}]{br15}
{del Palacio}, S., {Bosch-Ramon}, V., \& {Romero}, G.~E. 2015, \aap, 575, A112

\bibitem[{{Donat} {et~al.}(1998){Donat}, {Font}, {Ib{\'a}{\~n}ez}, \&
  {Marquina}}]{dfi98}
{Donat}, R., {Font}, J.~A., {Ib{\'a}{\~n}ez}, J.~M.~l., \& {Marquina}, A. 1998,
  Journal of Computational Physics, 146, 58

\bibitem[{{Donat} \& {Marquina}(1996)}]{DonMar96}
{Donat}, R. \& {Marquina}, A. 1996, Journal of Computational Physics, 125, 42

\bibitem[{{Dubus}(2006{\natexlab{a}})}]{dub06}
{Dubus}, G. 2006{\natexlab{a}}, \aap, 451, 9

\bibitem[{{Dubus}(2006{\natexlab{b}})}]{dub06*b}
{Dubus}, G. 2006{\natexlab{b}}, \aap, 456, 801

\bibitem[{{Dubus}(2013)}]{dub13}
{Dubus}, G. 2013, \aapr, 21, 64

\bibitem[{{Dubus} \& {Cerutti}(2013)}]{DubCer13}
{Dubus}, G. \& {Cerutti}, B. 2013, \aap, 557, A127

\bibitem[{{Dubus} {et~al.}(2008){Dubus}, {Cerutti}, \& {Henri}}]{dch08}
{Dubus}, G., {Cerutti}, B., \& {Henri}, G. 2008, \aap, 477, 691

\bibitem[{{Dubus} {et~al.}(2010){Dubus}, {Cerutti}, \& {Henri}}]{dch10}
{Dubus}, G., {Cerutti}, B., \& {Henri}, G. 2010, \aap, 516, A18

\bibitem[{{Dubus} {et~al.}(2015){Dubus}, {Lamberts}, \& {Fromang}}]{dlf15}
{Dubus}, G., {Lamberts}, A., \& {Fromang}, S. 2015, \aap, 581, A27

\bibitem[{{Font} {et~al.}(1994){Font}, {Ibanez}, {Marquina}, \&
  {Marti}}]{fim94}
{Font}, J.~A., {Ibanez}, J.~M., {Marquina}, A., \& {Marti}, J.~M. 1994, \aap,
  282, 304

\bibitem[{{Gould} \& {Schr{\'e}der}(1967)}]{GouSch67}
{Gould}, R.~J. \& {Schr{\'e}der}, G.~P. 1967, Physical Review, 155, 1404

\bibitem[{{Grove} {et~al.}(1995){Grove}, {Tavani}, {Purcell}, {Johnson},
  {Kurfess}, {Strickman}, \& {Arons}}]{gtp95}
{Grove}, J.~E., {Tavani}, M., {Purcell}, W.~R., {et~al.} 1995, \apjl, 447, L113

\bibitem[{{H.E.S.S.~Collaboration} {et~al.}(2013){H.E.S.S.~Collaboration},
  {Abramowski}, {Acero}, {Aharonian}, {Akhperjanian}, {Anton}, {Balenderan},
  {Balzer}, {Barnacka}, {Becherini}, {Becker Tjus}, {Bernl{\"o}hr}, {Birsin},
  {Biteau}, {Boisson}, {Bolmont}, {Bordas}, {Brucker}, {Brun}, {Brun}, {Bulik},
  {Carrigan}, {Casanova}, {Cerruti}, {Chadwick}, {Chaves}, {Cheesebrough},
  {Colafrancesco}, {Cologna}, {Conrad}, {Couturier}, {Dalton}, {Daniel},
  {Davids}, {Degrange}, {Deil}, {deWilt}, {Dickinson}, {Djannati-Ata{\"i}},
  {Domainko}, {Drury}, {Dubus}, {Dutson}, {Dyks}, {Dyrda}, {Egberts}, {Eger},
  {Espigat}, {Fallon}, {Farnier}, {Fegan}, {Feinstein}, {Fernandes},
  {Fernandez}, {Fiasson}, {Fontaine}, {F{\"o}rster}, {F{\"u}{\ss}ling},
  {Gajdus}, {Gallant}, {Garrigoux}, {Gast}, {Giebels}, {Glicenstein},
  {Gl{\"u}ck}, {G{\"o}ring}, {Grondin}, {Grudzi{\'n}ska}, {H{\"a}er}, {Hague},
  {Hahn}, {Hampf}, {Harris}, {Heinz}, {Heinzelmann}, {Henri}, {Hermann},
  {Hillert}, {Hinton}, {Hofmann}, {Hofverberg}, {Holler}, {Horns},
  {Jacholkowska}, {Jahn}, {Jamrozy}, {Jung}, {Kastendieck}, {Katarzy{\'n}ski},
  {Katz}, {Kaufmann}, {Kh{\'e}lifi}, {Klepser}, {Klochkov}, {Klu{\'z}niak},
  {Kneiske}, {Kolitzus}, {Komin}, {Kosack}, {Kossakowski}, {Krayzel},
  {Kr{\"u}ger}, {Lan}, {Lamanna}, {Lefaucheur}, {Lemoine-Goumard}, {Lenain},
  {Lennarz}, {Lohse}, {Lopatin}, {Lu}, {Marandon}, {Marcowith}, {Masbou},
  {Maurin}, {Maxted}, {Mayer}, {McComb}, {Medina}, {M{\'e}hault}, {Menzler},
  {Moderski}, {Mohamed}, {Moulin}, {Naumann}, {Naumann-Godo}, {de Naurois},
  {Nedbal}, {Nguyen}, {Niemiec}, {Nolan}, {Oakes}, {Ohm}, {de O{\~n}a
  Wilhelmi}, {Opitz}, {Ostrowski}, {Oya}, {Panter}, {Parsons}, {Paz Arribas},
  {Pekeur}, {Pelletier}, {Perez}, {Petrucci}, {Peyaud}, {Pita},
  {P{\"u}hlhofer}, {Punch}, {Quirrenbach}, {Raab}, {Raue}, {Reimer}, {Reimer},
  {Renaud}, {de los Reyes}, {Rieger}, {Ripken}, {Rob}, {Rosier-Lees}, {Rowell},
  {Rudak}, {Rulten}, {Sahakian}, {Sanchez}, {Santangelo}, {Schlickeiser},
  {Schulz}, {Schwanke}, {Schwarzburg}, {Schwemmer}, {Sheidaei}, {Skilton},
  {Sol}, {Spengler}, {Stawarz}, {Steenkamp}, {Stegmann}, {Stinzing}, {Stycz},
  {Sushch}, {Szostek}, {Tavernet}, {Terrier}, {Tluczykont}, {Trichard},
  {Valerius}, {van Eldik}, {Vasileiadis}, {Venter}, {Viana}, {Vincent},
  {V{\"o}lk}, {Volpe}, {Vorobiov}, {Vorster}, {Wagner}, {Ward}, {White},
  {Wierzcholska}, {Willmann}, {Wouters}, {Zacharias}, {Zajczyk}, {Zdziarski},
  {Zech}, \& {Zechlin}}]{haa13}
{H.E.S.S.~Collaboration}, {Abramowski}, A., {Acero}, F., {et~al.} 2013, \aap,
  551, A94

\bibitem[{{Johnston} {et~al.}(2005){Johnston}, {Ball}, {Wang}, \&
  {Manchester}}]{jbw05}
{Johnston}, S., {Ball}, L., {Wang}, N., \& {Manchester}, R.~N. 2005, \mnras,
  358, 1069

\bibitem[{{Johnston} {et~al.}(1992){Johnston}, {Manchester}, {Lyne}, {Bailes},
  {Kaspi}, {Qiao}, \& {D'Amico}}]{jml92}
{Johnston}, S., {Manchester}, R.~N., {Lyne}, A.~G., {et~al.} 1992, \apjl, 387,
  L37

\bibitem[{{Kennel} \& {Coroniti}(1984)}]{KenCor84}
{Kennel}, C.~F. \& {Coroniti}, F.~V. 1984, \apj, 283, 710

\bibitem[{{Khangulyan} \& {Aharonian}(2005)}]{KhaAha05}
{Khangulyan}, D. \& {Aharonian}, F. 2005, in American Institute of Physics
  Conference Series, Vol. 745, High Energy Gamma-Ray Astronomy, ed. F.~A.
  {Aharonian}, H.~J. {V{\"o}lk}, \& D.~{Horns}, 359--364

\bibitem[{{Khangulyan} {et~al.}(2008){Khangulyan}, {Aharonian}, \&
  {Bosch-Ramon}}]{kab08}
{Khangulyan}, D., {Aharonian}, F., \& {Bosch-Ramon}, V. 2008, \mnras, 383, 467

\bibitem[{{Khangulyan} {et~al.}(2011){Khangulyan}, {Aharonian}, {Bogovalov}, \&
  {Rib{\'o}}}]{kab11}
{Khangulyan}, D., {Aharonian}, F.~A., {Bogovalov}, S.~V., \& {Rib{\'o}}, M.
  2011, \apj, 742, 98

\bibitem[{{Khangulyan} {et~al.}(2012){Khangulyan}, {Aharonian}, {Bogovalov}, \&
  {Rib{\'o}}}]{kab12}
{Khangulyan}, D., {Aharonian}, F.~A., {Bogovalov}, S.~V., \& {Rib{\'o}}, M.
  2012, \apjl, 752, L17

\bibitem[{{Khangulyan} {et~al.}(2014{\natexlab{a}}){Khangulyan}, {Aharonian},
  \& {Kelner}}]{kak14}
{Khangulyan}, D., {Aharonian}, F.~A., \& {Kelner}, S.~R. 2014{\natexlab{a}},
  \apj, 783, 100

\bibitem[{{Khangulyan} {et~al.}(2014{\natexlab{b}}){Khangulyan}, {Bogovalov},
  \& {Aharonian}}]{kba14}
{Khangulyan}, D., {Bogovalov}, S.~V., \& {Aharonian}, F.~A. 2014{\natexlab{b}},
  International Journal of Modern Physics Conference Series, 28, 1460169

\bibitem[{{Khangulyan} {et~al.}(2007){Khangulyan}, {Hnatic}, {Aharonian}, \&
  {Bogovalov}}]{kha07}
{Khangulyan}, D., {Hnatic}, S., {Aharonian}, F., \& {Bogovalov}, S. 2007,
  \mnras, 380, 320

\bibitem[{{Kijak} {et~al.}(2011){Kijak}, {Dembska}, {Lewandowski}, {Melikidze},
  \& {Sendyk}}]{kdl11}
{Kijak}, J., {Dembska}, M., {Lewandowski}, W., {Melikidze}, G., \& {Sendyk}, M.
  2011, \mnras, 418, L114

\bibitem[{{Kishishita} {et~al.}(2009){Kishishita}, {Tanaka}, {Uchiyama}, \&
  {Takahashi}}]{ktu09}
{Kishishita}, T., {Tanaka}, T., {Uchiyama}, Y., \& {Takahashi}, T. 2009, \apjl,
  697, L1

\bibitem[{{Kong} {et~al.}(2012){Kong}, {Cheng}, \& {Huang}}]{kch12}
{Kong}, S.~W., {Cheng}, K.~S., \& {Huang}, Y.~F. 2012, \apj, 753, 127

\bibitem[{{Lamberts} {et~al.}(2013){Lamberts}, {Fromang}, {Dubus}, \&
  {Teyssier}}]{lfd13}
{Lamberts}, A., {Fromang}, S., {Dubus}, G., \& {Teyssier}, R. 2013, \aap, 560,
  A79

\bibitem[{{Lucy} \& {Solomon}(1970)}]{LucSol70}
{Lucy}, L.~B. \& {Solomon}, P.~M. 1970, \apj, 159, 879

\bibitem[{{Lyubarsky} \& {Kirk}(2001)}]{LyuKir01}
{Lyubarsky}, Y. \& {Kirk}, J.~G. 2001, \apj, 547, 437

\bibitem[{{Maraschi} \& {Treves}(1981)}]{MarTre81}
{Maraschi}, L. \& {Treves}, A. 1981, \mnras, 194, 1P

\bibitem[{{Mart{\'{\i}}} \& {M{\"u}ller}(2015)}]{MarMul15}
{Mart{\'{\i}}}, J.~M. \& {M{\"u}ller}, E. 2015, Living Reviews in Computational
  Astrophysics, 1

\bibitem[{{Mart{\'{\i}}} {et~al.}(1997){Mart{\'{\i}}}, {M{\"u}ller}, {Font},
  {Ib{\'a}{\~n}ez}, \& {Marquina}}]{mmf97}
{Mart{\'{\i}}}, J.~M., {M{\"u}ller}, E., {Font}, J.~A., {Ib{\'a}{\~n}ez},
  J.~M., \& {Marquina}, A. 1997, \apj, 479, 151

\bibitem[{{Moffat}(2008)}]{mof08}
{Moffat}, A.~F.~J. 2008, in Clumping in Hot-Star Winds, ed. W.-R. {Hamann},
  A.~{Feldmeier}, \& L.~M. {Oskinova}, 17

\bibitem[{{Morlino} {et~al.}(2015){Morlino}, {Lyutikov}, \& {Vorster}}]{mlv15}
{Morlino}, G., {Lyutikov}, M., \& {Vorster}, M. 2015, \mnras, 454, 3886

\bibitem[{{Negueruela} {et~al.}(2011){Negueruela}, {Rib{\'o}}, {Herrero},
  {Lorenzo}, {Khangulyan}, \& {Aharonian}}]{nrh11}
{Negueruela}, I., {Rib{\'o}}, M., {Herrero}, A., {et~al.} 2011, \apjl, 732, L11

\bibitem[{{Odell}(1981)}]{ode81}
{Odell}, S.~L. 1981, \apjl, 243, L147

\bibitem[{{Okazaki} {et~al.}(2011){Okazaki}, {Nagataki}, {Naito}, {Kawachi},
  {Hayasaki}, {Owocki}, \& {Takata}}]{onn11}
{Okazaki}, A.~T., {Nagataki}, S., {Naito}, T., {et~al.} 2011, \pasj, 63, 893

\bibitem[{{Olmi} {et~al.}(2014){Olmi}, {Del Zanna}, {Amato}, {Bandiera}, \&
  {Bucciantini}}]{oda14}
{Olmi}, B., {Del Zanna}, L., {Amato}, E., {Bandiera}, R., \& {Bucciantini}, N.
  2014, \mnras, 438, 1518

\bibitem[{{Paredes-Fortuny} {et~al.}(2015){Paredes-Fortuny}, {Bosch-Ramon},
  {Perucho}, \& {Rib{\'o}}}]{pbp15}
{Paredes-Fortuny}, X., {Bosch-Ramon}, V., {Perucho}, M., \& {Rib{\'o}}, M.
  2015, \aap, 574, A77

\bibitem[{{Perucho} {et~al.}(2004){Perucho}, {Hanasz}, {Mart{\'{\i}}}, \&
  {Sol}}]{phm04}
{Perucho}, M., {Hanasz}, M., {Mart{\'{\i}}}, J.~M., \& {Sol}, H. 2004, \aap,
  427, 415

\bibitem[{{Perucho} {et~al.}(2005){Perucho}, {Mart{\'{\i}}}, \&
  {Hanasz}}]{pmh05}
{Perucho}, M., {Mart{\'{\i}}}, J.~M., \& {Hanasz}, M. 2005, \aap, 443, 863

\bibitem[{{Porth} {et~al.}(2014){Porth}, {Komissarov}, \& {Keppens}}]{pkk14}
{Porth}, O., {Komissarov}, S.~S., \& {Keppens}, R. 2014, \mnras, 438, 278

\bibitem[{{Romoli} {et~al.}(2015){Romoli}, {Bordas}, {Mariaud}, {Murach},
  {Aharonian}, {de Naurois}, {P{\"u}hlhofer}, {Schwanke}, {van Soelen},
  {Sushch}, {Zabalza}, \& {for the H.~E.~S.~S.~Collaboration}}]{rbm15}
{Romoli}, C., {Bordas}, P., {Mariaud}, C., {et~al.} 2015, ArXiv e-prints

\bibitem[{{Shu} \& {Osher}(1988)}]{ShuOsh88}
{Shu}, C.-W. \& {Osher}, S. 1988, Journal of Computational Physics, 77, 439

\bibitem[{{Sierpowska} \& {Bednarek}(2005)}]{SieBed05}
{Sierpowska}, A. \& {Bednarek}, W. 2005, \mnras, 356, 711

\bibitem[{{Sierpowska-Bartosik} \& {Torres}(2007)}]{SieTor07}
{Sierpowska-Bartosik}, A. \& {Torres}, D.~F. 2007, \apjl, 671, L145

\bibitem[{{Sironi} \& {Spitkovsky}(2011)}]{SirSpi11}
{Sironi}, L. \& {Spitkovsky}, A. 2011, \apj, 741, 39

\bibitem[{{Takata} {et~al.}(2012){Takata}, {Okazaki}, {Nagataki}, {Naito},
  {Kawachi}, {Lee}, {Mori}, {Hayasaki}, {Yamaguchi}, \& {Owocki}}]{ton12}
{Takata}, J., {Okazaki}, A.~T., {Nagataki}, S., {et~al.} 2012, \apj, 750, 70

\bibitem[{{Tam} {et~al.}(2015){Tam}, {Li}, {Takata}, {Okazaki}, {Hui}, \&
  {Kong}}]{tlt15}
{Tam}, P.~H.~T., {Li}, K.~L., {Takata}, J., {et~al.} 2015, \apjl, 798, L26

\bibitem[{{Tavani} {et~al.}(1994){Tavani}, {Arons}, \& {Kaspi}}]{tak94}
{Tavani}, M., {Arons}, J., \& {Kaspi}, V.~M. 1994, \apjl, 433, L37

\bibitem[{{Uchiyama} {et~al.}(2009){Uchiyama}, {Tanaka}, {Takahashi}, {Mori},
  \& {Nakazawa}}]{utt09}
{Uchiyama}, Y., {Tanaka}, T., {Takahashi}, T., {Mori}, K., \& {Nakazawa}, K.
  2009, \apj, 698, 911

\bibitem[{{Volpi} {et~al.}(2008){Volpi}, {Del Zanna}, {Amato}, \&
  {Bucciantini}}]{vda08}
{Volpi}, D., {Del Zanna}, L., {Amato}, E., \& {Bucciantini}, N. 2008, \aap,
  485, 337

\bibitem[{{Yoon} \& {Heinz}(2016)}]{YooHei16}
{Yoon}, D. \& {Heinz}, S. 2016, ArXiv e-prints

\bibitem[{{Zabalza} {et~al.}(2013){Zabalza}, {Bosch-Ramon}, {Aharonian}, \&
  {Khangulyan}}]{zba13}
{Zabalza}, V., {Bosch-Ramon}, V., {Aharonian}, F., \& {Khangulyan}, D. 2013,
  \aap, 551, A17

\end{thebibliography}
\end{document}